\documentclass[prd,preprint,superscriptaddress,nofootinbib,titlepage]{revtex4-2}
\usepackage[english]{babel}
\usepackage{amsmath,amssymb,graphicx,xcolor}

\usepackage{dutchcal}
\usepackage{dsfont}
\usepackage{float}
\usepackage{hyperref}

\newcommand{\Tau}{\mathrm{T}}
\newcommand{\cdummy}{\cdot}
\newcommand{\nobracket}{}
\newcommand{\nosymbol}{}

\newcommand{\tmop}[1]{\ensuremath{\operatorname{#1}}}

\begin{document}
\count\footins = 1000

\title{Semiclassical gravitational collapse of a radially symmetric
massless scalar quantum field}

\author{Jana N. Guenther}
\email{jguenther@uni-wuppertal.de}
\affiliation{Aix Marseille Univ., Universit\'e de Toulon, CNRS, CPT, Marseille, France}
\affiliation{Department of Physics, University of Wuppertal, D-42119 Wuppertal, Germany}
\author{Christian Hoelbling}
\email{hch@uni-wuppertal.de}
\affiliation{Department of Physics, University of Wuppertal, D-42119 Wuppertal, Germany}
\author{Lukas Varnhorst}
\email{varnhorst@uni-wuppertal.de}
\affiliation{Aix Marseille Univ., Universit\'e de Toulon, CNRS, CPT, Marseille, France}
\affiliation{Department of Physics, University of Wuppertal, D-42119 Wuppertal, Germany}

\date{March 30, 2022}

\preprint{WUB/20-02}

\begin{abstract}
  We present a method to study the semiclassical gravitational collapse of a
  radially symmetric scalar quantum field in a coherent initial state. The
  formalism utilizes a Fock space basis in the initial metric, is unitary and
  time reversal invariant up to numerical precision. It maintains exact
  compatibility of the metric with the expectation values of the energy
  momentum tensor in the scalar field coherent state throughout the entire
  time evolution. We find a simple criterion for the smallness of
  discretization effects, which is violated when a horizon forms. As a first
  example, we study the collapse of a specific state in the angular
  momentum $l=0$ approximation. Outside the
  simulated volume it produces a Schwarzschild metric with $r_s \sim 3.5 \ell_p$. We see behaviour
  that is compatible with the onset of horizon formation both in the
  semiclassical and corresponding classical cases in a regime where we see no
  evidence for large discretization artefacts. In our example setting, we see
  that quantum effects accelerate the possible horizon formation and
  move it radially outward. We find that this effect is robust against
  variations of the radial resolution, the time step, the volume, the initial
  position and shape of the inmoving state, the vacuum subtraction, the
  discretization of the time evolution operator and the integration scheme of
  the metric. We briefly discuss potential improvements of the method and the
  possibility of applying it to black hole evaporation. We also
  briefly touch on the extension of our formalism to higher angular
  momenta, but leave the details and numerics for a forthcoming publication.
\end{abstract}

{\maketitle}

\section{Introduction}

Ever since Hawking's seminal paper {\cite{Hawking:1974sw}}
semiclassical effects in the vicinity of horizons have been widely debated in
the literature
\cite{Unruh:1976db,Jacobson:1995ab,tHooft:1984kcu,Almheiri:2012rt,Hooft:2016itl}
(for reviews see e.g. \cite{Jacobson:2003vx,Mathur:2009hf}). A lot of this
discussion has focused on the vacuum behaviour of quantum fields on a
background metric that is supposed to have formed by a different
mechanism than by the collapse of the quantum fields themselves.
In the classical theory, the formation of a
horizon from collapsing scalar fields has been intensively studied numerically
\cite{Choptuik:1992jv, Christodoulou:1986zr, Christodoulou:1986du,
  Christodoulou:1987vu, Christodoulou:1987vv, Goldwirth:1987nu,
  Roberts:1989sk, Brady:1994xfa,
  Oshiro:1994hd,Frolov:1998zt,Purrer:2004nq,Gundlach:2007gc},
while for the semiclassical case we are aware of only one other
suggestion for a numerical treatment
\cite{Berczi:2020nqy,Berczi:2021hdh}, although various other methods
have been applied to this problem \cite{Tomimatsu:1995qy, Bak:1999wb,
  Bak:2000kg, Vachaspati:2018pps,Kawai:2020rmt}.

In this paper we present a formalism to numerically study the gravitational
collapse of a radially symmetric complex scalar quantum field in real
time in the angular momentum $l=0$ approximation. 

The fact that the scalar field only interacts semiclassically via gravity
allows us to trace its evolution without any Fock space truncation if we
remain in the Fock space of the original metric. While it would be extremely
complicated to represent the scalar field in the Fock space of the final
metric and thus observe the outgoing particle content, we can nonetheless
trace the expectation values of crucial quantities such as the Hamiltonian
density of the scalar field throughout the entire time evolution and thereby
obtain a quantitative picture of the collapse.

One of the main features of our formalism is the guaranteed compatibility of
the scalar field with the background metric at all times. Although we can
disentangle classical from vacuum contributions and thus explicitly study
backreaction effects, the metric is at every point in our time evolution
compatible with the corresponding expectation value of the energy-momentum
tensor of the scalar field.

\begin{figure}[th!]
  \begin{center}
    \includegraphics[width=0.7\textwidth]{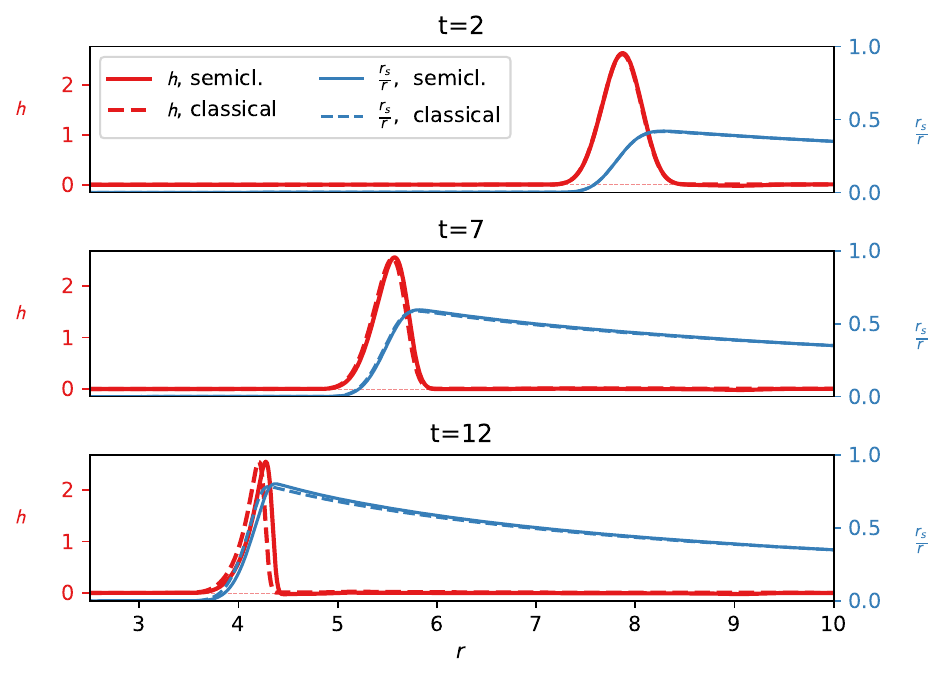}
  \end{center}
  \caption{\label{result}Three snapshots of the gravitational collapse of a
  spherically symmetric scalar field. The plots show, for three different
  asymptotic observer times, the radial Hamiltonian density of the scalar
  field (or its expectation value in the case of the quantum field) and the
  ratio of local Schwarzschild radius to radius $r_s / r$ for both the
  classical and semiclassical case. It is clearly visible
  that quantum effects increase the peak of the forming horizon as well as
  shifting its location radially outward.}
\end{figure}

Using this new formalism, we then investigate the collapse of one specific
field configuration that asymptotically gives a Schwarzschild metric outside of the simulated volume with (in
Planck units) $r_s \sim 3.5$. We demonstrate that with very moderate
computational effort our formalism can trace the time evolution of such a
state faithfully until about a maximum $r_s / r \sim 0.9$ signifying the onset
of horizon formation. We investigate the sizes of discretization and finite
volume effects and apply different vacuum subtraction procedures,
discretizations and integration schemes. In addition, we give a simple
criterion that signifies the presence of large discretization artefacts. In
our example setting, the onset of the horizon formation in the semiclassical
theory happens at larger radii and with a higher $r_s / r$ than in the
classical case (see fig.~\ref{result}). This finding is corroborated by an
energy influx into the region of the forming horizon both from the outside and
from the inside.

Continuing our simulations at times even closer to the expected horizont
formation, we see a host of new effects that we deem likely to be
discretization artefacts.

Although it is well known that $l=0$ modes dominate the Hawking
radiation in the spherically symmetric case \cite{Page:1976df}, and
that the $l=0$ approximation, which we employ here, can be used to
obtain some qualitative insight \cite{Callan:1992rs}, we would like to
stress that it is still an uncontrolled approximation in principle,
and that taking higher momentum modes into account will be essential
for a quantitative understanding. The formalism as presented in this
paper is not directly suited for the complete numerical treatment of
the system, including the vacuum contributions to the higher angular
momentum modes of to the scalar field.  A generalisation taking these
modes into account is possible \cite{Guenther:2021wkd} and will be
studied in detailed in a forthcoming publication.

The paper is structured as follows. In sect. \ref{deriv} we derive the
semiclassical (and classical) equations of motion for coherent states in our
original Fock space basis. In sect. \ref{numimp} we discretize these equations
of motion and provide a method for implementing sensible consistent initial
states. In section \ref{res} we present the main physics results and we
conclude in sect. \ref{conc}.

\section{Derivation of the formalism}\label{deriv}

\subsection{The classical equations of motions}

We work in Planck units $\hbar = c = G = 1$. We investigate a radially
symmetric classical $N_c$
component complex scalar field $\overline{\phi}$ on a background metric of the
form {\cite{Choptuik:1992jv}}
\begin{equation}
  g_{\mu \nu} = \mathrm{diag} (\alpha^2 (t, r), - a^2 (t, r), - r^2, - r^2
  \sin^2 \theta) 
\end{equation}
From the action
\[
  S=\int \mathrm{d}^4x\sqrt{-g}\left(
    -\frac{1}{16\pi}R
    +\frac{1}{2}g^{\mu\nu}\overline{\phi}^{\dag}_{,\mu}\overline{\phi}_{,\mu}
    \right)
\]
we obtain a complete set of classical
equations of motion \cite{Choptuik:1992jv, Gundlach:2007gc}. 
For the metric part we have
\begin{equation}
  \begin{split}
    \frac{\alpha_{, r}}{\alpha} - \frac{a_{, r}}{a} - \frac{a^2 - 1}{r} & = 
    0\\
    \frac{a_{, r}}{a} + \frac{a^2 - 1}{2 r} & = \frac{a}{r \nosymbol \alpha}
    \mathcal{H}_r\\
    \frac{a_{, t}}{a} & = \frac{\alpha}{r \nosymbol a} \mathcal{P}_r
  \end{split} \label{eomstart}
\end{equation}
while  the evolution of the scalar field is governed by the Hamiltonian
$\mathcal{H}= \int_0^{\infty} \mathrm{d} r\mathcal{H}_r$ arising from the
Hamiltonian density
\[ \mathcal{H}_r = \frac{\alpha}{a} \left( \frac{1}{2 \pi r^2} \overline{\Pi}
   \overline{\Pi}^{\dag} + 2 \pi r^2 \overline{\phi}^{\dag}_{, r}
   \overline{\phi}_{, r} \right) \]
where the conjugate momenta are defined as
\[ \overline{\Pi} = 2 \pi r^2 \frac{a}{\alpha} \overline{\phi}^{\dag}_{, t}
   \qquad \overline{\Pi}^{\dag} = 2 \pi r^2 \frac{a}{\alpha}
   \overline{\phi}_{, t} \]
In addition, a pseudo-momentum density
\[ \mathcal{P}_r = (\overline{\Pi} \overline{\phi}_{, r} +
   \overline{\Pi}^{\dag} \overline{\phi}^{\dag}_{, r}) \]
occurs in the last equation in (\ref{eomstart}). This equation however is
not independent and we ignore it in our numerical treatment other than for
crosschecking purposes.

We now perform the substitution
\[ \phi = \sqrt{2\pi} r \sqrt{\frac{a^0}{\alpha^0}} \overline{\phi} \]
so that the Hamiltonian density may be written as
\begin{equation}
  \mathcal{H}_r = (Q \psi)^{\dagger}_r  (Q \psi)_r
\end{equation}
where
\begin{equation}
  Q = \left(\begin{array}{cc}
    q & 0\\
    0 & A
  \end{array}\right) \qquad \psi = \left(\begin{array}{c}
    \phi\\
    \Pi^{\dagger}
  \end{array}\right)
\end{equation}
with the diagonal matrix $A$ that has elements
\begin{equation}
  A_r = \frac{a^0_r \alpha_r}{\alpha^0_r a_r} \label{adef}
\end{equation}
The metric parameters $a^0$ and $\alpha^0$ here refer to the values of $a$ and
$\alpha$ at a reference time $t_0$ and the operator $q$ is defined as
\begin{equation}
  q = \sqrt{\frac{\alpha}{a}} r \partial_r \sqrt{\frac{\alpha^0}{a^0}} 
  \frac{1}{r} \label{qdef}
\end{equation}
To facilitate a mode decomposition of the field $\phi$, we introduce the
singular value decomposition (SVD) of $q_0$, the operator $q$ at time $t_0$,
as
\begin{equation}
  q^0 = U \omega V^T, \ \ \omega = \mathrm{diag} (\omega_1, \omega_2, \ldots) .
  \label{svdq0}
\end{equation}
This decomposition allows us to define field modes $\hat{\psi}_k$, which
depend on the metric at the reference time $t_0$, as
\begin{equation}
  \psi_r = \int_0^{\infty} \mathrm{d} k \hspace{0.17em} U^{\ast}_{rk} 
  \hat{\psi}_k
\end{equation}
Here and in the following, quantities with an index $k$ are understood to be
related to the modes defined in the above equation. We adopted an index
notation for $k$ and $r$ even for the continuous case. Using the basis
\[ \begin{split}
     \hat{\phi}_k = \frac{(b_-)_k + (b_+)^{\dagger}_k}{\sqrt{2 \omega}} &\qquad
     \hat{\phi}_k^{\dagger} = \frac{(b_-)^{\dagger}_k + (b_+)_k}{\sqrt{2
     \omega}}\\
     \hat{\Pi}_k = - \frac{(b_-)_k - (b_+)^{\dagger}_k}{\sqrt{2 \omega}} &\qquad
     \hat{\Pi}_k^{\dagger} = \frac{(b_-)^{\dagger}_k - (b_+)_k}{\sqrt{2
     \omega}}
   \end{split} \]
we express the Hamiltonian density as
\[ \begin{split}
     \mathcal{H}_r= & \frac{1}{2}  \int_0^{\infty} \mathrm{d} k' 
     \int_0^{\infty} \mathrm{d} k \hspace{0.17em} \sqrt{\omega_k \omega_{k'}}
     A_r\\
     &   (((b_+)_{k'} (b_+)^{\dagger}_k + (b_-)^{\dagger}_{k'} (b_-)_k)\\
     &   (U_{rk'} U_{rk} + V_{rk'} V_{rk}) +\\
     &   ((b_+)_{k'} (b_-)_k + (b_-)^{\dagger}_{k'} (b_+)^{\dagger}_k)\\
     &   (U_{rk'} U_{rk} - V_{rk'} V_{rk}))
   \end{split} \]
The poisson brackets of the $(b_{\pm})_k$ and $(b_{\pm})^{\dagger}_k$ are
\begin{equation}
  \{(b_+)_k, (b_+)^{\dagger}_{k'} \} = \{(b_-)_k, (b_-)^{\dagger}_{k'} \} = -
  i \delta (k - k') \mathds{1}_f
\end{equation}
where $\mathds{1}_f$ is the unit matrix in component space and all other
Poisson brackets vanish. In the semiclassical theory, the $(b_{\pm})_k^{\dag}$
and $(b_{\pm})_k$ will therefore play the role of creation and annihilation
operators of the scalar field.

\subsection{Time evolution of observables}

In anticipation of the semiclassical time evolution, we will now develop the
equations of motion for the $(b_{\pm})_k^{\dag}$ and $(b_{\pm})_k$, keeping in
mind that they can be either complex numbers obeying the canonical equations
of motion or operators with commutation relations
\begin{equation}
  [(b_+)_k, (b_+)^{\dagger}_{k'}] = [(b_-)_k, (b_-)^{\dagger}_{k'}] = \delta
  (k - k') \mathds{1}_f
\end{equation}
We start by writing the Hamiltonian in the compact form
\[ \mathcal{H}= b_+ W b_+^{\dag} + b_-^{\dag} W b_- + b_+ X b_- + b_-^{\dag} X
   b_+^{\dag} \]
where we have defined
\[ W_{k' k} = \int_0^{\infty} \mathrm{d} r \frac{1}{2} \sqrt{\omega_{k'}}
   \left( U_{r k'}  \frac{a_0 \alpha}{\alpha_0 a} U_{r k} + V_{r k'} 
   \frac{a_0 \alpha}{\alpha_0 a} V_{r k} \right) \sqrt{\omega_k} \]
and
\[ X_{k' k} = \int_0^{\infty} \mathrm{d} r \frac{1}{2} \sqrt{\omega_{k'}}
   \left( U_{r k'}  \frac{a_0 \alpha}{\alpha_0 a} U_{r k} - V_{r k'} 
   \frac{a_0 \alpha}{\alpha_0 a} V_{r k} \right) \sqrt{\omega_k} \]
The Heisenberg equations of the $(b_{\pm})_k$ and
$(b_{\pm})^{\dagger}_k$ on a given metric then become\footnote{Here
  we introduce an integral convention $x_k y_k := \int_0^{\infty}
  \mathrm{d} k  x_k y_k$ over continuous momentum modes. For discrete
  indices, the standard summation convention still applies.}
\begin{equation}
  \begin{split}
    (b_+)_{k, t} & = - i (b_+)_{k'} W_{k' k} - i (b_-)^{\dag}_{k'} X_{k'
    k}\\
    (b_-)_{k, t} & = - i W_{k k'} (b_-)_{k'} - i X_{k k'} (b_+)^{\dag}_{k'}
  \end{split} \label{heq1}
\end{equation}
which can easily be shown to be a Bogolyubov {\cite{Bogolyubov:1958km}}
transformation in the semiclassical case. We would now like to construct the
creation and annihilation operators at a given time $b_{\pm}^{\dag} (t)$ and
$b_{\pm} (t)$ in terms of the same operators \ $b_{\pm}^{\dag} =
b_{\pm}^{\dag} (t_0)$ and $b_{\pm} = b_{\pm} (t_0)$ at the reference time
$t_0$. Because of the form of the time evolution (\ref{heq1}) we can
generically write
\[ \begin{split}
     b_+ (t) & = b_+ w_+ (t) + b_-^{\dag} x_+^{\dag} (t)\\
     b_- (t) & = w_- (t) b_- + x_-^{\dag} (t) b_+^{\dag}
   \end{split} \]
with coefficient matrices $w_{\pm} (t)$ and $x_{\pm} (t)$ that obviously
fulfill the initial conditions
\[ w_{\pm} (t_0) =\mathds{1} \qquad x_{\pm} (t_0) = 0 \]
Using these initial conditions and the time evolution of the creation- and
annihilation operators (\ref{heq1}) we find the time evolution of the
coefficient matrices to be
\footnote{We use the shorthand notation $\dot{x}:=x_{,t}$.}
\[ \begin{split}
     \dot{w} (t) & = - i w (t) W - i x^{\ast} (t) X\\
     \dot{x} (t) & =- i x (t) W - i w^{\ast} (t) X
   \end{split} \]
where we can identify
\[ w (t) = w_+ (t) = w_-^T (t) \qquad x (t) = x_+^{\dag} (t) = x_-^{\ast} (t)
\]
Since the time evolution is a Bogolyubov transformation, the coefficient
matrices fulfill the identities
\[ \mathds{1}= w^{\dag} (t) w (t) - x^{\dag} (t) x (t) \qquad x^T (t) w (t) =
   w^T (t) x (t) \]
We now define linear combinations
\[ \begin{split}
     u (t) & = (w (t) + x^{\ast} (t)) \sqrt{\omega} U^T\\
     v (t) & = (w (t) - x^{\ast} (t)) \sqrt{\omega} V^T
   \end{split} \]
of the coefficient matrices for which the time evolution equations
\begin{equation}
  \begin{split}
    \dot{u} & = - i v A q^{0 T}\\
    \dot{v} & = - i u A q^0
  \end{split} \label{eomuv}
\end{equation}
are particularly well suited for a numerical treatment. We can express
$\mathcal{H}_r$ and $\mathcal{P}_r$ at an arbitrary time $t$ in terms of these
new coefficient matrices and the creation/annihilation operators at the
initial time $t_0$ as
\begin{equation}
  \begin{split}
    \mathcal{H}_r = & \frac{1}{2} (b_+ v + b_-^{\dag} v^{\ast}) A_r
    (v^{\dag} b_+^{\dag} + v^T b_-)\\
     & + \frac{1}{2} (b_+ u - b_-^{\dag} u^{\ast}) A_r (u^{\dag} b_+^{\dag}
    - u^T b_-)
  \end{split} \label{Hdef}
\end{equation}
and
\begin{equation}
  \begin{split}
    \mathcal{P}_r = & \frac{i \nosymbol}{2} (b_+ u + b_-^{\dag} u^{\ast})
    \frac{a^0_r }{\alpha^0_r } (v^{\dag} b_+^{\dag} - v^T b_-)\\
    & - \frac{i \nosymbol}{2} (b_+ v - b_-^{\dag} v^{\ast})
    \frac{a^0_r}{\alpha^0_r} (u^{\dag} b_+^{\dag} + u^T b_-)
  \end{split} \label{Pdef}
\end{equation}
The first two equations of (\ref{eomstart}) together with (\ref{adef},
\ref{qdef}, \ref{eomuv}, \ref{Hdef}, \ref{Pdef}) and the initial conditions
\begin{equation}
  \begin{split}
    u (t_0) = & \sqrt{\omega} U^T\\
    v (t_0) = & \sqrt{\omega} V^T
  \end{split} \label{iniuv}
\end{equation}
thus form a complete set of equations of motion for the classical case. As a
last step, we replace the parameters $\alpha$ and $a$ describing the metric by
the more suitable
\[ d = \frac{r}{a^2} \qquad \hat{\alpha} = \alpha a \]
which we can use to rewrite the equations of motion (\ref{eomstart}) as
\begin{equation}
  \begin{split}
    \frac{1 - d_{, r}}{d} & = \frac{\hat{\alpha}_{, r}}{\hat{\alpha}}\\
    \frac{1 - d_{, r}}{2 d} & =  \hat{\mathcal{H}}_r\\
    - \frac{1}{2 \hat{\alpha}} \frac{d_{, t}}{d} & = \hat{\mathcal{P}}_r
  \end{split} \label{eomcl2}
\end{equation}
with
\begin{equation}
  \begin{split}
    \hat{\mathcal{H}}_r = & \frac{1}{2} (b_+ v + b_-^{\dag} v^{\ast})
    \frac{1}{\hat{\alpha}^0_r d^0_r } (v^{\dag} b_+^{\dag} + v^T b_-)\\
     & + \frac{1}{2} (b_+ u - b_-^{\dag} u^{\ast})
    \frac{1}{\hat{\alpha}^0_r d^0_r } (u^{\dag} b_+^{\dag} - u^T b_-)
  \end{split} \label{ham2}
\end{equation}
and
\begin{equation}
  \begin{split}
    \hat{\mathcal{P}}_r  = & \frac{i \nosymbol}{2} (b_+ u + b_-^{\dag}
    u^{\ast}) \frac{1}{\hat{\alpha}^0_r d^0_r } (v^{\dag} b_+^{\dag} - v^T
    b_-)\\
     & - \frac{i \nosymbol}{2} (b_+ v - b_-^{\dag} v^{\ast})
    \frac{1}{\hat{\alpha}^0_r d^0_r } (u^{\dag} b_+^{\dag} + u^T b_-)
  \end{split} \label{p2}
\end{equation}
The relation to the original densities is given by
\begin{equation}
  \hat{\mathcal{H}}_r = \frac{1}{\hat{\alpha}_r d_r } \mathcal{H}_r \qquad
  \hat{\mathcal{P}}_r = \frac{1}{\hat{\alpha}_r d_r } \mathcal{P}_r
  \label{hatrel}
\end{equation}
We also recast the auxiliary variable (\ref{adef}) as
\[ A_r = \frac{\hat{\alpha}_r d_r}{\hat{\alpha}^0_r d^0_r} \]
and the operator $q^0 = q (t_0)$ (\ref{qdef})
\begin{equation}
  q^0 = \sqrt{r \hat{\alpha}^0 d^0} \partial_r \sqrt{\frac{\hat{\alpha}^0
  d^0}{r^3}} \label{defq0}
\end{equation}
Note that $\hat{\mathcal{H}}_r$ does not depend on the current metric at all,
but only on the metric at the reference time $t_0$. It is thus easy to
radially integrate the first two equations in (\ref{eomcl2}).

\subsection{Vacuum subtraction and normal ordering}

We now proceed to quantize the scalar field in the Heisenberg picture. For
this purpose we can utilize the standard Fock space representation at the
reference time $t_0$, since all the subsequent time evolution is absorbed by
the coefficient matrices $u$ and $v$. We also need to replace the
$\mathcal{H}_r$ and $\mathcal{P}_r$ in (\ref{eomstart}) by expectation values
of suitably normal ordered operators.

Let us start by considering a general bilinear operator
\[
     O = o^0 + b_+ o^{+ +} b_+^{\dag} + b_-^{\dag} o^{- -} b_- + b_+ o^{+
     -} b_- + b_-^{\dag} o^{- +} b_+^{\dag}
   \]
Its vacuum expectation value is
\[ \langle 0 | O | 0 \rangle = o^0 + \tmop{Tr} (o^{+ +}) \]
and thus if we want it to vanish, we have to impose the normal ordering
condition
\[ o^0 = - \tmop{Tr} (o^{+ +}) \]
If we add to the modified Hamiltonian density (\ref{ham2}) a constant term
$\hat{h}^0_r$ so that
\[ : \hat{\mathcal{H}}_r : = \hat{h}^0_r + \hat{\mathcal{H}}_r \]
the normal ordering condition becomes
\[ \hat{h}^0_r = - \frac{N_c}{2} \frac{1}{\hat{\alpha}^0_r d^0_r }
   (v^{\ast}_{k r} v_{k r} + u^{\ast}_{k r} u_{k r}) \]
Naturally, this condition can not be satisfied for all times simultaneously.
Given the physical situation we are interested in, however, we choose the following
two subtraction schemes. First, we simply demand that at our reference time $t_0$, where
the non-zero energy density is far away from a potentially forming horizon, the
vacuum expectation value of the Hamiltonian density vanishes. This results in
\begin{equation}
  \hat{h}^0_r = - \frac{N_c}{2} \frac{1}{\hat{\alpha}^0_r d^0_r } (v^{\ast}_{k
  r} (t_0) v_{k r} (t_0) + u^{\ast}_{k r} (t_0) u_{k r} (t_0)) \label{no1}
\end{equation}
where the initial $u_{k r} (t_0)$ and $v_{} (t_0)$ are given by (\ref{iniuv}).
This fails to exactly reproduce the desired property, namely that the vacuum
expectation value of the Hamiltonian density vanishes on a flat background
metric, but it approaches it asymptotically as we shift the initial position
of our inmoving field further outward. The second possibility is to
try to directly render the Hamiltonian density zero on a flat background
metric. To accomplish this, we can in principle define
\begin{equation}
  \hat{h}^0_r = - \frac{N_c}{2} (v^{\tmop{free} \ast}_{k r} v^{\tmop{free}}_{k
  r} + u^{\tmop{free} \ast}_{k r} u^{\tmop{free}}_{k r}) \label{no2}
\end{equation}
with the free
\[ \begin{split}
     u^{\tmop{free}} = & \sqrt{\omega_{\tmop{free}}} U^T_{\tmop{free}}\\
     v^{\tmop{free}} = & \sqrt{\omega_{\tmop{free}}} V^T_{\tmop{free}}
   \end{split} \]
where $U_{\tmop{free}}$, $V_{\tmop{free}}$ and $\omega_{\tmop{free}}$ are
obtained by an SVD of the free operator
\[ U_{\tmop{free}} \omega_{\tmop{free}} V^T_{\tmop{free}} = q_{\tmop{free}} =
   r \partial_r  \frac{1}{r} \]
In this paper, we predominantly use the first definition, but we will also
explore the possibility of imposing the second condition in our numerical
simulations.
Several other subtraction schemes are possible and it would be interesting to study if they lead to different behaviours. We leave this question
for future work.
We note in passing that $\langle 0 | \hat{\mathcal{P}} (t_0) | 0
\rangle = 0$ and thus no vacuum subtraction needs to be applied for this operator.

There is a third type of vacuum subtraction that is interesting purely
as an observable. We call it the final state vacuum subtraction and it
consists of subtracting the vacuum hamiltonian density
\begin{equation}
  \hat{h}^0_r = - \frac{N_c}{2}\frac{1}{\hat{\alpha}_r d_r } (\overline{v}^{\ast}_{k r} \overline{v}_{k
  r} + \overline{u}^{\ast}_{k r} \overline{u}_{k r}) \label{nof}
\end{equation}
where the
\[ \begin{split}
     \overline{u} = & \sqrt{\overline{\omega}} \overline{U}^T\\
     \overline{v} = & \sqrt{\overline{\omega}} \overline{V}^T
   \end{split} \]
 belong to the operator
 \begin{equation}
  \overline{q} = \sqrt{r \hat{\alpha} d} \partial_r \sqrt{\frac{\hat{\alpha}
  d}{r^3}} \label{qbar}
\end{equation}
that has an SVD
\begin{equation}
  \overline{q} = \overline{U} \overline{\omega} \overline{V}^{\Tau} \label{qbsvd}
\end{equation}
It corresponds to subtracting the vacuum Hamiltonian density of the
vacuum state corresponding to the metric parameters
$\hat{\alpha}=\hat{\alpha}(t)$ and $d=d(t)$ that have evolved at a
time $t$. Since it is extremely difficult to reexpress states other
than the vacuum in this new Fock basis, we will not be using it for
the time evolution of our system, but it is nonetheless an interesting
observable.

\subsection{coherent states}

Let us now construct a single particle state
\[ | f \rangle = (f_+^{\dag} b_+^{\dag} + b_-^{\dag} f_-) | 0 \rangle \]
where the $f_{\pm i k}$ are as of now arbitrary coefficients of the
$i^{\tmop{th}}$ field component corresponding to the SVD component (frequency)
$\omega_k$. We will use the shorthand notation
\[ b^{\dag} = (f_+^{\dag} b_+^{\dag} + b_-^{\dag} f_-) \]
and for convenience impose unit normalization on the state
\[ \langle f | \nobracket f \rangle = f_+^{\dag} f_+ + f_-^{\dag} f_- = 1 \]
Taking the expectation value of $: \hat{\mathcal{H}}_r :$ in this state
results in
\[ \begin{split}
     \langle f | : \hat{\mathcal{H}}_r : | f \rangle = & \frac{N_c}{2}
     \frac{1}{\hat{\alpha}^0_r d^0_r } (v^{\ast}_{k r} v_{k r} + u^{\ast}_{k
     r} u_{k r}) + \hat{h}^0_r\\
      & + \frac{1}{2} \left( f_+^T \left( u^{\ast}
     \frac{1}{\hat{\alpha}^0_r d^0_r } u^T + v^{\ast}
     \frac{1}{\hat{\alpha}^0_r d^0_r } v^T \right) f_+^{\ast} \right)\\
      & + \frac{1}{2} \left( f_-^{\dag} \left( u^{\ast}
     \frac{1}{\hat{\alpha}^0_r d^0_r } u^T + v^{\ast}
     \frac{1}{\hat{\alpha}^0_r d^0_r } v^T \right) f_- \right)
   \end{split} \]
while
\[ \begin{split}
     \langle f | : \hat{\mathcal{P}}_r : | f \rangle = &
     \frac{N_c}{\hat{\alpha}^0_r d^0_r } \tmop{Im} (v^{\ast}_{k r} u_{k r})\\
      & + \frac{i}{2} \left( f_+^T \left( v^{\ast} 
     \frac{1}{\hat{\alpha}^0_r d^0_r } u^T - u^{\ast} 
     \frac{1}{\hat{\alpha}^0_r d^0_r} v^T \right) f_+^{\ast} \right)\\
      & + \frac{i}{2} \left( f_-^{\dag} \left( v^{\ast}
     \frac{1}{\hat{\alpha}^0_r d^0_r } u^T - u^{\ast}
     \frac{1}{\hat{\alpha}^0_r d^0_r } v^T \right) f_- \right)
   \end{split} \]
It is straightforward to generalize these relations for arbitrary $n$-particle
states
\[ | n \rangle = \frac{1}{\sqrt{n!}} b^{\dag n} | 0 \rangle \]
We will however concentrate on unit normalized coherent states
\[ | \lambda \rangle = e^{- \frac{| \lambda |^2}{2}} \sum_{n = 0}^{\infty}
   \frac{\lambda^n}{\sqrt{n!}} | n \rangle \]
with the coherent state parameter $\lambda$. The relevant matrix elements of
coherent states are given by
\[ \begin{split}
     \langle \lambda \nobracket | b_{k +}^{\dag} b_{k' +} | \nobracket \lambda
     \rangle = & | \lambda |^2 f_+ (k) f_+^{\ast} (k')\\
     \langle \lambda \nobracket | b_{k -}^{\dag} b_{k' -} | \nobracket \lambda
     \rangle = & | \lambda |^2 f_-^{\ast} (k) f_- (k')\\
     \langle \lambda \nobracket | b_{k' +} b_{k -} | \nobracket \lambda
     \rangle = & \lambda^2 f_- (k) f^{\ast}_+ (k')\\
     \langle \lambda \nobracket | b_{k' -}^{\dag} b_{k +}^{\dag} | \nobracket
     \lambda \rangle = & \lambda^{\ast 2} f_-^{\ast} (k') f_+ (k)
   \end{split} \]
If we thus define
\[ l_{k +} = \lambda^{\ast} f_+ (k) \hspace{3em} l_{k -} = \lambda f_- (k) \]
we can cast the matrix elements $\hat{\mathcal{h}}_r = \langle \lambda | :
\hat{\mathcal{H}}_r : | \lambda \rangle$ and $\hat{\mathcal{p}}_r = \langle
\lambda | \hat{\mathcal{P}}_r | \lambda \rangle$ into the form
\begin{equation}
  \begin{split}
    \hat{\mathcal{h}}_r = & \frac{N_c}{2 \hat{\alpha}^0_r d^0_r }
    (v^{\ast}_{k r} v_{k r} + u^{\ast}_{k r} u_{k r}) + \hat{h}^0_r\\
     & + \frac{1}{2 \hat{\alpha}^0_r d^0_r } ((l_u)^{\dag}_r (l_u)_r +
    (l_v)^{\dag}_r (l_v)_r)
  \end{split} \label{hamcoh}
\end{equation}
\begin{equation}
  \begin{split}
    \hat{\mathcal{p}}_r= & \frac{N_c}{\hat{\alpha}^0_r d^0_r } \tmop{Im}
    (v^{\ast}_{k r} u_{k r})\\
     & + \frac{1}{2 \hat{\alpha}^0_r d^0_r } ((l_u)^{\dag}_r (l_v)_r +
    (l_v)^{\dag}_r (l_u)_r)
  \end{split} \label{pcoh}
\end{equation}
where
\begin{equation}
  \begin{split}
    l_u = & - i (u^{\dag} l_+ + u^T l_-)\\
    l_v = & v^{\dag} l_+ - v^T l_-
  \end{split} \label{ludef}
\end{equation}
Note that the second line on the r.h.s of both (\ref{hamcoh}) and (\ref{pcoh})
are equivalent to the right hand sides of (\ref{ham2}) resp. (\ref{p2}) when
substituting $(l^{\dag}_{+},l_{-})$ for $b_{\pm}$ and thus correspond to the classical
component. The first lines on the r.h.s. of (\ref{hamcoh}) and (\ref{pcoh})
thus represent the quantum effects. Instead of the classical equations of
motion (\ref{eomcl2}), we now have the semiclassical equations of motion
\begin{equation}
  \begin{split}
    \frac{1 - d_{, r}}{d} = & \frac{\hat{\alpha}'}{\hat{\alpha}}\\
    \frac{1 - d_{, r}}{2 d} = &  \hat{\mathcal{h}}_r\\
    - \frac{1}{2 \hat{\alpha}} \frac{d_{, t}}{d} = & \hat{\mathcal{p}}_r
  \end{split} \label{eoms}
\end{equation}
It is straightforward to show that the third equation is redundant also in the
semiclassical case with coherent states and that the entire set of equations
is self consistent. We also find that the quantities $\hat{\mathcal{h}}$ and
$\hat{\mathcal{p}}$ fulfill a modified continuity equation
\begin{equation}
  \frac{1}{a} \left( \frac{a}{\alpha} \mathcal{h} \right)_{, t} -
  \frac{1}{\alpha} \left( \frac{\alpha}{a} \mathcal{p} \right)_{, r} = 0
  \label{conteq}
\end{equation}
where, in accordance with (\ref{hatrel}) we have
\begin{equation}
  \mathcal{h}= \hat{\alpha} d \hat{\mathcal{h}} \qquad \mathcal{p}=
  \hat{\alpha} d \hat{\mathcal{p}} \label{hhatdef}
\end{equation}
so that $\mathcal{p}$ gives the direction of the scalar field energy flux with
the convention that a positive $\mathcal{p}$ corresponds to an inward pointing
flux.

We will from now on exclusively focus on states that excite only a single
field component, e.g. the first one. We can thus drop the field index
component and the only effect of the additional field components is the
enhancement by $N_c$ of the vacuum contribution. Following \cite{Kawai:2020rmt} we use this as a tool to enhance vacuum effects in case one
needs them to be more prominent.

\section{Numerical implementation}\label{numimp}

The first two equations of (\ref{eoms}) together with (\ref{hamcoh},
\ref{ludef}, \ref{eomuv}, \ref{defq0}), the initial conditions (\ref{iniuv})
and the SVD (\ref{svdq0}) form a complete set of equations to describe the
time evolution of our field that we now want to discretize. As noted in the
preceding section, we can effectively use the same set of equations to
describe both the classical and the semiclassical evolution with the only
difference that in the classical evolution we have to omit the first lines on
the r.h.s. of (\ref{hamcoh}). Our numerical integration scheme is based on
leap-frogging the scalar evolution equations (\ref{eomuv}) with a radial
integration of the metric according to the first two equations of (\ref{eoms})
in such a way as to optimally conserve the Bogolyubov-transformation property
of the time evolution.

\subsection{The classical radial integration}

Let us first concentrate on the classical radial integration of the metric
given $\hat{\mathcal{h}}_r$. We introduce a set of radial coordinates $r_i$,
$i = 0, \ldots, N_r$ with $\Delta_i = r_i - r_{i - 1} > 0$ and the boundary
conditions of our metric $r_0 = d_0$ and $\hat{\alpha}_p = \hat{\alpha} (r_p)
= 1$. The latter of the two conditions ensures that our metric is
Schwarzschild for $r > r_{N_r}$ (with a Schwarzschild radius $r_p - d_p$)
while the former places a horizon at $r_0$ if $r_0 > 0$ or ensures that there
is no central singularity at $r = 0$ if $r_0 = 0$. In the numerical part of
this paper, we will exclusively study the case $r_0 = 0$ although the
algorithm presented is perfectly capable of numerically handling the $r_0 > 0$
case as well.

To integrate the metric, we note that the first two equations of (\ref{eoms})
give us
\[ \ln (\hat{\alpha})' = 2 \hat{\mathcal{h}} \]
which integrates to
\[ \hat{\alpha}_{i - 1} = \hat{\alpha}_i e^{- 2 h_i} \]
where we have defined
\[ h_i = \int_{r_{i - 1}}^{r_i} d \nosymbol r \hat{\mathcal{h}} (r) \]
Together with the boundary condition $\hat{\alpha}_{N_r} = 1$ we thus find
\begin{equation}
  \hat{\alpha}_i = e^{- 2 \sum_{j = i + 1}^p h_j} \label{intalhat}
\end{equation}
To obtain $d_i = d (r_i)$ we have to integrate the second equations in
(\ref{eoms}), i.e.
\[ \frac{1 - d'}{2 d} = \hat{\mathcal{h}}_r \]
To do so, we must assume a specific shape of the density function in between
two discretization points and varying this shape will result in different
discretization errors. For the numerical simulations we choose either of two
shapes, both of which are safe in a sense that they do not produce unwanted
horizons ($d (r) = 0$) in between discretization points. The first shape we
use is a series of $\delta$-shells
\[ \hat{\mathcal{h}} (r) = \lim_{\varepsilon \rightarrow 0}
   \left\{\begin{array}{lll}
     \frac{h_i}{\varepsilon} &  & r_i - \varepsilon < r \leqslant r_i\\
     0 &  & \tmop{else}
   \end{array}\right. \]
which results in the recursion relation
\begin{equation}
  d_i = (d_{i - 1} + \Delta_i) e^{- 2 h_i} \label{intd1}
\end{equation}
The second shape is a piecewise constant function
\[ \hat{\mathcal{h}} (r) = \frac{h_i}{r_i - r_{i - 1}} \hspace{3em} r_{i - 1}
   < r < r_i \]
which results in the recursion relation\footnote{Note that in the numerical
implementation one can use Horner's method to obtain accurate values of $d_i$
for small $h_i$. Written in Horner form, the Taylor expansion of the critical
function is
\[ \frac{\sinh (x)}{x} = 1 + \frac{x^2}{2 \cdummy 3} \left( 1 +
   \frac{x^2}{4 \cdummy 5} \left( 1 + \frac{x^2}{6 \cdummy 7} (1 + \ldots) \right)
   \right) \]
For $x < 0.3$ it is sufficient to use this expansion up to the fifth
nontrivial term to guarantee double precision accuracy.}
\begin{equation}
  d_i = e^{- h_i} \left( d_{i - 1} e^{- h_i} + \Delta_i \frac{\sinh
  (h_i)}{h_i} \right) \label{intd2}
\end{equation}
Either of these two relations allows us to compute the $d_i$ from inside out
starting with the boundary condition $d_0 = 0$. We can also see that the
relative difference between both recursion relations is of the order $h_i$ and
thus it is important that the condition
\begin{equation}
  h_i \ll 1 \label{contcond}
\end{equation}
is maintained throughout the entire time evolution. This is not trivial
especially near a forming horizon. From (\ref{hhatdef}) we conclude that for a
finite $\mathcal{h}$, $\hat{\mathcal{h}}$ diverges when $d \rightarrow 0$.
Since a horizon is characterized by $d = 0$, our formalism will inevitably
break down at some point before a horizon appears.

\subsection{Construction of a consistent initial state}

Before we start the time evolution of the scalar field, we need to ensure that
we have a consistent initial state, i.e. the initial coherent state
coefficients $l_{k \pm}$ \ when inserted into (\ref{hamcoh}) have to produce
an $\hat{\mathcal{h}}$ that generates the correct initial metric parameters
$\hat{\alpha}^0$ and $d^0$. These parameters occur in the operator $q^0$
(\ref{defq0}) whose SVD (\ref{svdq0}) in turn provides the basis (\ref{iniuv})
for the mode expansion of the initial state. To construct such a self
consistent initial state we start by providing initial $h_i \geqslant 0$ and
the ratio $k_i \in [- 1, 1]$. We proceed by constructing from them the initial
metric parameters according to (\ref{intalhat}) and either (\ref{intd1}) or
(\ref{intd2}). This allows us to construct the operator $q^0$ according to
(\ref{defq0}) and obtain its SVD (\ref{svdq0}). Of course we have to
discretize the derivative operator $\partial_r$ in $q_0$ and we do this by
taking the simple forward difference operator. Other options and their
consequences are explored in sect. \ref{numimp}.

The next step is to determine the classical part of the initial $h_i$ which
we denote by $h_i^c = h_i - h_i^0$ where $h_i^0$ is the vacuum contribution.
When imposing the normal ordering condition (\ref{no1}), the initial vacuum
expectation value $\langle 0 | : \hat{\mathcal{H}}_r : | 0 \rangle = 0$, so
$h_i^0 = 0$ and thus $h_i^c = h_i$. For the normal ordering condition
(\ref{no2}) we instead have
\begin{equation}
  \begin{split}
    h_i^0 = & \frac{N_c}{2} \frac{1}{\hat{\alpha}^0_i d^0_i } (v^{\ast}_{k
    i} (t_0) v_{k i} (t_0) + u^{\ast}_{k i} (t_0) u_{k i} (t_0))\\
     & - \frac{N_c}{2} (v^{\tmop{free} \ast}_{k i} v^{\tmop{free}}_{k i} +
    u^{\tmop{free} \ast}_{k i} u^{\tmop{free}}_{k i})
  \end{split} \label{vss}
\end{equation}
It turns out that, especially in regions where $h_i = 0$, we can have a vacuum
contribution that is slightly larger than the total $h^0_i > h_i$. This would
imply a negative classical contribution $h_i^c < 0$ which of course is
impossible. To obtain an approximate free vacuum subtraction we thus have to
modify the initial vacuum contribution slightly. Since it turns out that the
overall shape of the vacuum contribution $h^0_i$ is rather similar to $h_i$
itself up to a multiplicative factor, our preferred choice for an approximate
free vacuum subtraction is thus provided by
\[ h^c_i = (1 - x) h_i \]
where the constant $x$ is given by the ratio of the vacuum contribution
\[ x = \frac{h_m^0}{h_m} \]
at the position $m$ of the maximum value of $h_i .$

With the classical contribution $h^c_i$ thus obtained we construct the
\[ \begin{split}
     (l_u)_i = & \tmop{sign} (k_i) \sqrt{h^c_i \left( 1 \mp \sqrt{1 - k_i^2}
     \right)}\\
     (l_v)_i = & \sqrt{h_i^c \left( 1 \pm \sqrt{1 - k_i^2} \right)}
   \end{split} \]
and, inverting relation (\ref{ludef}) for the initial values $u (t_0)$ and $v
(t_0)$ (\ref{iniuv}), we finally obtain the consistent initial state
\begin{equation}
    l_{k \pm} = \frac{1}{\sqrt{\omega_k}} \sum_{i = 1}^p \frac{\pm (l_v)_i
    V_{i k} + i (l_u)_i U_{i k}}{2}
  \label{inil}
\end{equation}

\subsection{Initializing the scalar field}

The physical situation we are interested in is the gravitational collapse of a
scalar field. Accordingly, we would like the initial state to be an inmoving
shell. To make it inmoving as much as possible, we choose $k_i = 1$, which
results in $p_i (t_0) = h_i (t_0)$ where the $h_i$ and $p_i$ are the
discretized versions of (\ref{hamcoh}) and (\ref{pcoh})
\begin{equation}
  \begin{split}
    h_i = & \frac{N_c}{2 \hat{\alpha}^0_i d^0_i } (v^{\ast}_{k i} v_{k i} +
    u^{\ast}_{k i} u_{k i}) + \hat{h}^0_i\\
     & + \frac{1}{2 \hat{\alpha}^0_i d^0_i } ((l_u)^{\dag}_i (l_u)_i +
    (l_v)^{\dag}_i (l_v)_i)\\
     & \\
    p_i = & \frac{N_c}{\hat{\alpha}^0_i d^0_i } \tmop{Im} (v^{\ast}_{k i}
    u_{k i})\\
     & + \frac{1}{2 \hat{\alpha}^0_i d^0_i } ((l_u)^{\dag}_i (l_v)_i +
    (l_v)^{\dag}_i (l_u)_i)\\
     & \\
    l_u = & - i (u^{\dag} (t) l_+ + u^T (t) l_-)\\
    l_v = & v^{\dag} (t) l_+ - v^T (t) l_-
  \end{split} \label{hamfin}
\end{equation}
We are still left with the choice of the exact shape of $h_i$. One could in
principle take a thin shell at some radius $r_e$, i.e. $h_i \propto \delta_{i
e} / \Delta_e$, but this or similar choices have a lot of high frequency (i.e.
large $\omega$) components that we would like to avoid. We therefore choose to
smear out the thin shell to a bump that has a finite support $[R - \sigma, R +
\sigma]$. When studying discretization artefacts, we choose
the width of the bump to stay constant in physical units. We therefore
initialize
\begin{equation}
  h_i (t_0) = \left\{\begin{array}{lll}
    \Delta_i f_{\sigma, \lambda} (r_i - R) &  & | R - r_i | < \sigma\\
    0 &  & | R - r_i | \geqslant \sigma
  \end{array}\right. \label{bumpshape}
\end{equation}
In this paper, we use either one of two window functions. The first one, which
we use for our main results, is the Nuttall bump function {\cite{1163506}}
that has the form
\begin{equation}
  f_{\sigma, \lambda} (x) = \lambda \sum_{k = 0}^3 a_k \cos \left( k \frac{2
  \pi (x + \sigma)}{\sigma} \right) \label{nut}
\end{equation}
with the coefficients
\[ \begin{split}
     a_0 = & 0.355768\\
     a_1 = & - 0.487396\\
     a_2  = & 0.144232\\
     a_3  = & - 0.012604
   \end{split} \]
and a normalization factor $\lambda$ that determines the bump height. The
second, which we use as a crosscheck, is
\begin{equation}
  f_{\sigma, \lambda} (x) = \lambda e^{- \frac{\sigma^2}{\sigma^2 - x^2}}
  \label{bmp}
\end{equation}
We prefer the Nuttall function as it excites high frequency components (in
$\omega_k$) less. Since the Nuttall function is designed to suppress high frequency
components in Fourier space as opposed to our $U$-$V$ basis, further
improvement that is targeted to the specific basis should be possible. We do
however not investigate this in the current paper but leave it for future
studies.

\subsection{Time evolution of the scalar field}\label{core}

Our next task is to evolve the scalar field component matrices $u$ and $v$
according to (\ref{eomuv}) with the initial condition (\ref{iniuv}). The
update should be as close as possible to an exact Bogolyubov transformation.
We start out by noting that we can write (\ref{eomuv}) in the form
\[ \begin{split}
     \dot{u} \sqrt{\frac{\hat{\alpha} d}{\hat{\alpha}^0 d^0}} = & - i v
     \sqrt{\frac{\hat{\alpha} d}{\hat{\alpha}^0 d^0}} \overline{q}^T\\
     \dot{v} \sqrt{\frac{\hat{\alpha} d}{\hat{\alpha}^0 d^0}} = & - i u
     \sqrt{\frac{\hat{\alpha} d}{\hat{\alpha}^0 d^0}} \overline{q}
   \end{split} \]
with the operator (\ref{qbar})
that has an SVD (\ref{qbsvd}).
We define
\[ g = \sqrt{\frac{\hat{\alpha} d}{\hat{\alpha}^0 d^0}} \qquad H = g
   \left(\begin{array}{cc}
     0 & \overline{U} \overline{\omega} \overline{V}^{\Tau}\\
     \overline{V} \overline{\omega} \overline{U}^T & 0
   \end{array}\right) g^{- 1} \]
and write the time evolution in matrix form as
\[ \left(\begin{array}{ccc}
     \mathrm{d}u &  & \mathrm{d}v
   \end{array}\right) = - i \left(\begin{array}{ccc}
     u &  & v
   \end{array}\right) H\mathrm{d}t \]
Assuming constant metric parameters and thus constant $g$, we can integrate
this expression over a finite time intervall $\Delta t$ and obtain
\[ \left(\begin{array}{ccc}
     u (t + \Delta t) &  & v (t + \Delta t)
   \end{array}\right) = \left(\begin{array}{ccc}
     u (t) &  & v (t)
   \end{array}\right) e^{- i H \Delta t} \]
Computing the matrix exponential we find
\[ e^{- i H \Delta t} = g \left(\begin{array}{cc}
     \overline{U} \cos (\overline{\omega} \Delta t) \overline{U}^{\Tau} & - i
     \overline{U} \sin (\overline{\omega} \Delta t) \overline{V}^{\Tau}\\
     - i \overline{V} \sin (\overline{\omega} \Delta t) \overline{U}^T & \overline{V}
     \cos (\overline{\omega} \Delta t) \overline{V}^{\Tau}
   \end{array}\right) g^{- 1} \]
A constant metric during the time evolution of the scalar field is of course an
approximation that introduces finite time step discretization errors. In order
to keep these small, we implemented an update scheme that leap-frogs the time
evolution of the scalar field with the radial integration of the metric. In
particular, we start out with an implicit time step that updates $u$ and $v$
from $t - \Delta t$ to the time $t$ according to
\begin{equation}
  \begin{split}
    u (t) = & u (t - \Delta t) g \overline{U} \cos (\overline{\omega} \Delta t)
    \overline{} \overline{} \overline{U}^{\Tau} g^{- 1}\\
     & - i v (t - \Delta t) g \overline{V} \sin (\overline{\omega} \Delta t)
    \overline{} \overline{U}^{\Tau} g^{- 1}\\
     & \\
    v (t) = & v (t - \Delta t) g \overline{V} \cos (\overline{\omega} \Delta t)
    \overline{} \overline{} \overline{V}^{\Tau} g^{- 1}\\
     & - i u (t - \Delta t) g \overline{U} \sin (\overline{\omega} \Delta t)
    \overline{} \overline{V}^{\Tau} g^{- 1}
  \end{split} \label{uvimp}
\end{equation}
where the metric parameters $d$ and $\hat{\alpha}$ that occur in $g$ are those
at time $t$. To obtain these metric parameters, we start by first evaluating
(\ref{uvimp}) with the metric parameters at $t - \Delta t$. We then start an
iteration by first computing $h_i$ according to (\ref{hamfin}). This allows us
to compute updated metric parameters $d$ and $\hat{\alpha}$ according to
(\ref{intalhat}) and either (\ref{intd1}) or (\ref{intd2}) that we can plug in
to $\overline{q}$ (\ref{qbar}) and $g$. Following an other SVD (\ref{qbsvd}),
we can complete one iteration step by reevaluating (\ref{uvimp}). This
iteration is repeated until the metric parameters converge, which is typically
achieved after a few steps for cases of interest. After having successfully
iterated the metric parameters $d$ and $\hat{\alpha}$ and thus $g$ at time
$t$, we complete the update by performing an explicit time step
\[ \begin{split}
     u (t + \Delta t) = & u (t) g \overline{U} \cos (\overline{\omega} \Delta t)
     \overline{} \overline{} \overline{U}^{\Tau} g^{- 1}\\
      & - i v (t) g \overline{V} \sin (\overline{\omega} \Delta t) \overline{}
     \overline{U}^{\Tau} g^{- 1}\\
      & \\
     v (t + \Delta t) = & v (t) g \overline{V} \cos (\overline{\omega} \Delta t)
     \overline{} \overline{} \overline{V}^{\Tau} g^{- 1}\\
      & - i u (t) g \overline{U} \sin (\overline{\omega} \Delta t) \overline{}
     \overline{V}^{\Tau} g^{- 1}
   \end{split} \]
with them. One can show that this update is an exact Bogolyubov
transformation, which in the component matrix basis is equivalent to
the condition
\[
  \tmop{Re} (u^{\dag}(t) v(t))=q^0
\]
being fulfilled for all times $t$.

\subsection{Absorbing boundary conditions\label{absorbing_bc}}

The update algorithm detailed thus far is unitary and has, apart from
numerical errors, an exact time reversal symmetry. A wave packet that hits the
outer boundary $r_p$ of our discretized space will be reflected. Of course
this is a finite volume effect and can in principal be remedied by expanding
the range of our discretization. However, if we are willing to give up
unitarity, we can implement absorbing boundary conditions, for the classical
part at least, by modifying the initial state at every time step. Let us for
this purpose define the classical part of $h_i$ and $p_i$ from (\ref{hamfin})
\[ \begin{split}
     h_i^c = & \frac{1}{2 \hat{\alpha}^0_i d^0_i } ((l_u)^{\dag}_i (l_u)_i +
     (l_v)^{\dag}_i (l_v)_i)\\
     p_i^c = & \frac{1}{2 \hat{\alpha}^0_i d^0_i } ((l_u)^{\dag}_i (l_v)_i +
     (l_v)^{\dag}_i (l_u)_i)
   \end{split} \]
and impose the condition that both $h^c$ and $p^c$ vanish on the outermost
coordinate $r_{N_r}$, i.e. $h^c_{N_r} = p^c_{N_r} = 0$. This condition implies
\[ (l_u)_i = (l_v)_i = 0 \]
which, according to the relation (\ref{ludef}) requires a change in the
coherent state coefficients $l_{\pm}$. According to (\ref{inil}), these
coefficients have the form
\begin{equation}
  l_{\pm} = \pm l_R + i l_I \label{ldec}
\end{equation}
with real $l_R$ and $l_I$ which we want to preserve. We thus plug the ansatz
(\ref{ldec}) into (\ref{ludef}) and solve for $l_R$ and $l_I$. The result is
\[ \begin{split}
     l_R = & \frac{1}{2} (v_R^T + v_I^T u_R^{T - 1} u_I^T)^{- 1} (l_v -
     v_I^T u_R^{T - 1} l_u)\\
     l_I = & \frac{1}{2} (u_R^T + u_I^T v_R^{T - 1} v_I^T)^{- 1} (l_u +
     u_I^T v_R^{T - 1} l_v)
   \end{split} \]
where $u_{R / I}$ and $v_{R / I}$ are the real resp. imaginary parts of $u$
and $v$. Obviously this modification of the state is not part of the time
evolution in the Heisenberg picture. We could in principle model the absorbed
component as an outgoing classical wave packet in some approximation for $r >
r_{N_r}$ and feed the resulting $\hat{\alpha}_{N_r} < 1$ back into the radial
integration of the metric as a boundary condition. For the sake of simplicity
however, we restrict ourselves to maintaining $\hat{\alpha}_{N_r} = 1$ which
implies that the absorbed component vanishes in a truly unphysical way.

\subsection{Numerical stability}

Having implemented the algorithm we described above numerically in double
precision arithmetic, we encountered problems with numerical stability of the
SVD in certain cases. Specifically, if $r_0 = 0$ and the $r_i$ are evenly
spaced the numerical inaccuracies in in the vacuum component of the $h_1$
increase exponentially once we have $N_r \gtrsim 200$ discretization points.
Of course one can solve this problem by increasing the numerical precision,
which is however extremely expensive in terms of computer time. Fortunately
there is an alternative for the specific physical situation that we are
interested in: Initially (i.e. at time $t_0$), we choose the scalar field configuration as a
thick shell that has vanishing $\hat{\mathcal{h}} (r) = \hat{\mathcal{p}} (r)
= 0$ for $r < R_0$. With the specific window functions we have defined in
(\ref{bumpshape}), we have $R_0 = R - \sigma$. We can thus integrate a the
trajectory $r_{\tmop{cut}} (t)$ of a lightlike test particle that is radially
free falling towards the center starting from $R_0$ at $t_0$. If we exclude
causality violations (which can only appear as discretization effects), we
know that at the time $t$ we should have vanishing $\hat{\mathcal{h}} (r) =
\hat{\mathcal{p}} (r) = 0$ inside a shell $r < r_{\tmop{cut}} (t)$. Imposing
this condition during the update, i.e. explicitly setting to zero all $h_i$
for which $r_i < r_{\tmop{cut}} (t)$, will thus remove these particular
discretization artefacts.

In order to implement this procedure, we need to integrate the trajectory of
our test particle. A radially ingoing lightlike geodesic in our metric is
characterized by
\[ \mathrm{d}t = - \frac{\hat{\alpha} d}{r} \mathrm{d}r \]
In a region where $\hat{\mathcal{h}} = 0$ (which applies to the entire region
that our test particle is free falling through), $\hat{\alpha}$ and $\rho = r
- d$ are constant and the geodesic integrates to
\[ (t_1 - t_2) \hat{\alpha} = r_2 - r_1 + \rho \ln \frac{r_2 - \rho}{r_1 -
   \rho} \]
Solving explicitly for $r_2$ we obtain
\[ r_2 = \rho + W_0 ((r_1 - \rho) e^{r_1 - \rho + (t_1 - t_2) \hat{\alpha}})
\]
where $W_0 (x)$ is the principal branch of the Lambert $W$-function. When
performing a complete update step of our scalar field and the metric by a time
intervall $\Delta t$, we thus need to update the cutoff radius as
\[ r_{\tmop{cut}} (t + \Delta t) = \rho + W_0 ((r_{\tmop{cut}} (r) - \rho)
   e^{r_{\tmop{cut}} (t) - \rho + \Delta t \hat{\alpha}}) \]
\section{Results}\label{res}

\subsection{Simulation setup}

We will now concentrate on one specific physical situation which is that of an
$N_c = 2$ component field that is in a geometry with no preexisting horizon,
i.e. $r_0 = 0$. We will also choose a uniform discretization in the radial
coordinate $r$, i.e. $r_i - r_{i - 1} = \Delta$ for all $i = 1, \ldots, N_r$.
Although non-uniform discretizations would be desirable and are in principle
possible with our formalism, we observed partial reflection phenomena at the
boundary between different $\Delta_i$ whose further investigation we leave
open for future studies. We vary the physical size $r_{N_r}$ of our
discretized system to be either $10$, $12$ or $14$ and at $t_0 = 0$ place
initially inmoving bumps of the generic form (\ref{bumpshape}) with width
$\sigma = 1$ at either $R = 9, 11$ or $13$ with the obvious restriction $R <
r_{N_r}$ so that the bump fits into the discretized system. The size of the
initial bump is chosen such that the effective Schwarzschild radius of the
outermost shell is $\rho_{N_r} = r_{N_r} - d_{N_r} \cong 3.5$. We have
different radial discretizations in steps of factors 2 between $\Delta = 0.1$
and $\Delta = 0.00625$ corresponding, for a system size of $r_{N_r} = 10$, to
$N_r = 100, 200, 400, 800$ and $1600$ and time steps $\Delta t = 0.001, 0.002$
and $0.004$. In the following discussion, we will use as a default case the
system with $r_{N_r} = 10$ ($R = 9$), $N_r = 800$, $\Delta t = 0.004$, a
Nuttall bump shape (\ref{nut}), a simple forward difference discretization of
the derivative operator in (\ref{defq0}, \ref{qbar}), an initial state vacuum
subtraction (\ref{no1}) and a $\delta$-shell radial integration (\ref{intd1})
with reflecting boundary conditions.

Note that on a technical level the time evolution of the classical
field is equivalent to setting $N_c = 0$ in (\ref{hamfin}). When we
refer to the vacuum and classical contributions to any observable, we
generically mean the coefficient of the term proportional to $N_c$ and
the rest. We can thus refer to the ``vacuum contribution'' even in the
classical ($N_c = 0$) case, where it does not contribute towards the
time evolution at all. Thus defined, we can directly compare the
vacuum contributions of the classical and semiclassical time
evolution, which is useful for estimating backreaction effects.

Throughout the time evolution of our system, we will monitor how well the
condition (\ref{contcond}) is fulfilled, which is necessary for radial
discretization effects to be small. For this reason we will define a safe zone
through the criterion\footnote{The specific value 0.08 used in this criterion
applies to the Nuttall bump shape and may vary slightly for other shapes.}
\begin{equation}
  \max_i (h_i) = \max (\Delta \hat{\mathcal{h}}) < 0.08 \label{safezone}
\end{equation}
that we will elaborate on in sect. \ref{contlim}. For our default case, this
translates into a safe zone of $t \lesssim 14$.8 with values for other
discretizations listed in tab. \ref{tsafe}.

\begin{table}[th!]
  \begin{center}
  \begin{tabular}{r|l}
    $N_r$ & $t_{\tmop{safe}}$\\
    \hline
    100 & 4.8\\
    200 & 8.9\\
    400 & 12.1\\
    800 & 14.8\\
    1600 & 17.5
  \end{tabular}
\end{center}
  \caption{\label{tsafe}The maximum time for which the criterion
  (\ref{safezone}) is fulfilled for our semiclassical default case and its
  variations due to different radial discretization.}
\end{table}

Note that this criterion is a necessary but not a sufficient condition for the smallness of discretization artefacts.

\subsection{Qualitative behaviour of the system}

\begin{figure}[th!]
  \begin{center}
    \includegraphics[width=0.7\textwidth]{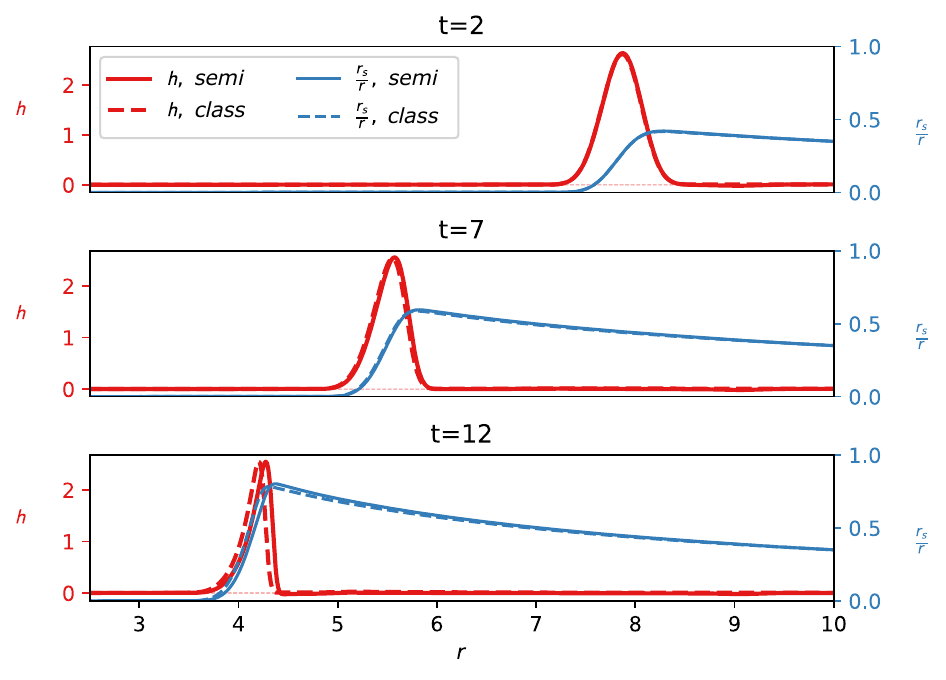}
\end{center}
\caption{\label{def1}Time evolution of the Hamiltonian density $\mathcal{h}$
  and the local $r_s / r$ for our default run. We can clearly see the onset of
  horizon formation. In the semiclassical case the horizon formation is both
  more pronounced and happening at larger $r$.}
\end{figure}

\begin{figure}[th!]
  \begin{center}
    \includegraphics[width=0.7\textwidth]{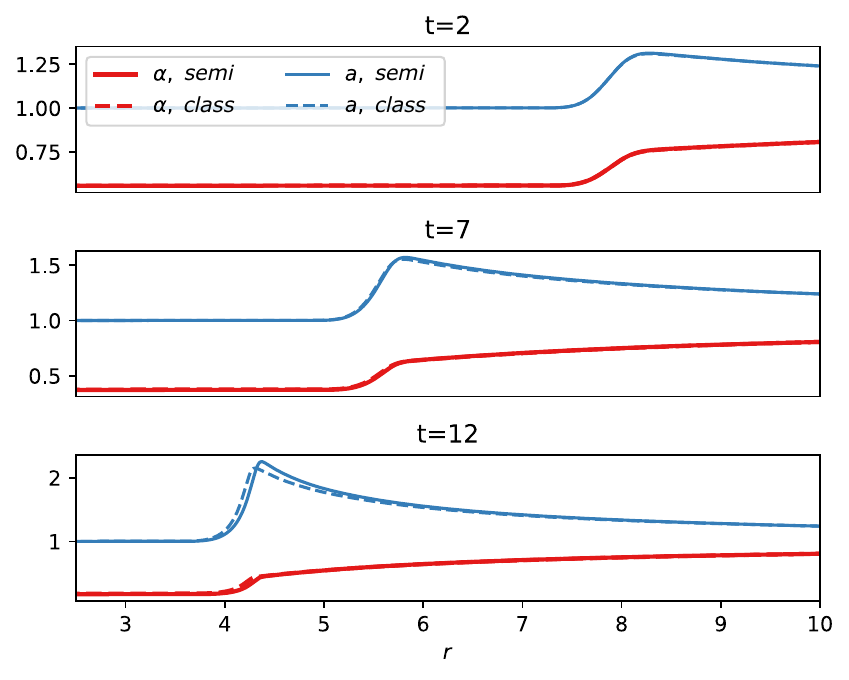}
\end{center}
\caption{\label{defmet}Time evolution of the metric parameters $\alpha$ and
  $a$ for our default run. The sharper and more radially outward onset of
  horizon formation in the semiclassical case is clearly visible.}
\end{figure}

Let us first look at the time evolution of our default case in the safe zone.
In fig.~\ref{def1} we compare at three different (asymptotic coordinate) times
$t$ the semiclassical ($N_c = 2$) to the classical ($N_c = 0$) time
evolution. We plot the Hamiltonian density\footnote{For simplicity we will
routinely omit to mention that we are looking at expectation values in the
semiclassical case and assume that this is implied.} $\mathcal{h}=
\hat{\alpha} d h$ together with the metric parameter $r_s / r = 1 - d / r$
that indicates how far away a system is at any given point from forming a
horizon. The behaviour of both the classical and the semiclassical case is
consistent with a forming horizon. We note however, that quantum effects both
enhance the peak in $r_s / r$ and shift it radially outward. This behaviour is
also evident when looking at the original metric parameters $\alpha$ and $a$
(fig.~\ref{defmet}). \

\begin{figure}[th!]
  \begin{center}
    \includegraphics[width=0.7\textwidth]{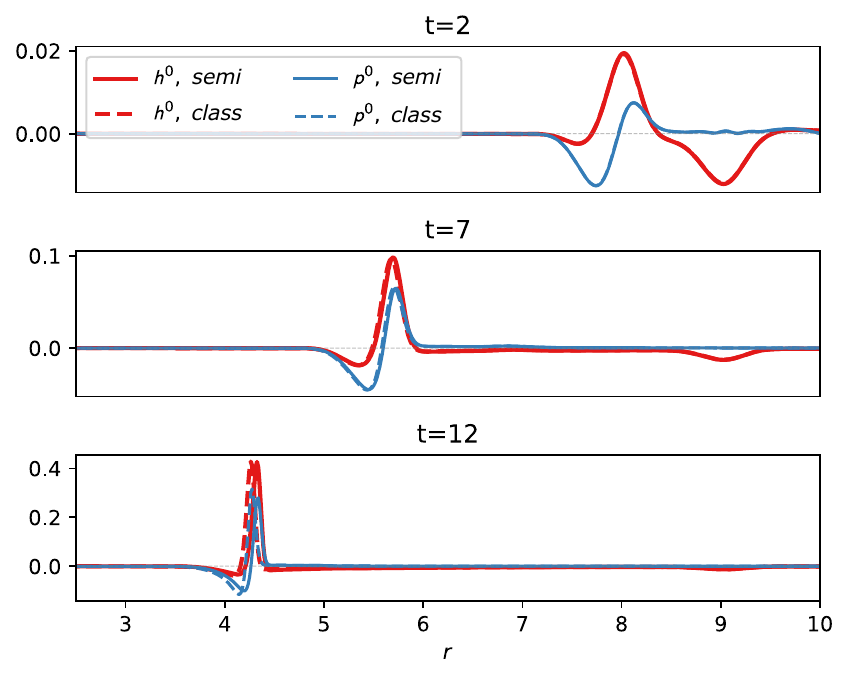}
\end{center}

  \
  \caption{\label{defvac}Time evolution of the vacuum contributions to the
  densities $\mathcal{h}$ and $\mathcal{p}$. For all times shown, the shape of
  $\mathcal{p}^0$ consists of a negative bump interior to the forming horizon
  and a positive bump outside of it, indicating an additional energy influx
  towards the horizon forming region from both the inside and the outside.
  Correspondingly, the vacuum contribution to the energy density
  $\mathcal{h}^0$ is positive in the horizon forming region and negative just
  outside it. The dip in $\mathcal{h}^0$ at the original position of the bump
  $r = 9$ is an effect of the vacuum subtraction procedure.}
\end{figure}

\begin{figure}[th!]
  \begin{center}
    \includegraphics[width=0.7\textwidth]{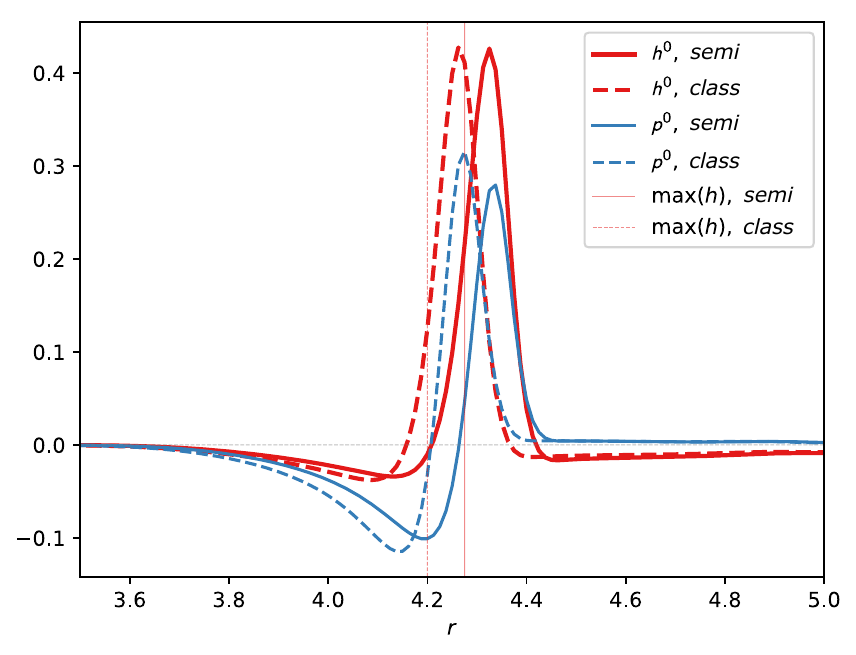}
  \end{center}
  \caption{\label{detailpeak}A more detailed look at the relevant region of
  the $t = 12$ panel of fig.~\ref{defvac}. Horizontal lines indicate the peak
  position of the total hamiltonian density $\mathcal{h}$. It is evident that
  quantum effects increase the peak height of $\mathcal{h}$ and shifting it
  outward while slightly decreasing its value in the immediate vicinity.}
\end{figure}

To identify the origin of this quantum effect, we plot in fig.~\ref{defvac}
the vacuum contribution to the hamiltonian density $\mathcal{h}^0$ together
with the vacuum contribution $\mathcal{p}^0$ of $\mathcal{p}$, which is
related to $\mathcal{h}$ by a modified continuity equation (\ref{conteq}) and
thus akin to a momentum density. We see that for all three time slices plotted
in fig.~\ref{defvac}, $\mathcal{p}^0$ has the structure of a leading negative
region followed by a trailing positive one. The zero crossing happens close to
the point of maximum $\mathcal{h}$ (see fig.~\ref{detailpeak}).
Remembering that in our convention a positive value of $\mathcal{p}$ signifies
a radial influx while a negative value corresponds to an outflux, the shape of
$\mathcal{p}^0$ tells us that quantum effects tend to increase the peak in the
hamiltonian density $\mathcal{h}$ at the expense of the neighbouring
regions in this phase of the collapse.
This is clearly visible in the shape of $\mathcal{h}^0$, which has a peak
flanked on either side by regions where it is negative. It is also interesting
to note that this structure is even visible in the classical case with the
peak of $\mathcal{h}^0$ shifted accordingly, so that it is slightly outward of
the peak of $\mathcal{h}$ as it is in the semiclassical case. In fig.
\ref{detailpeak} we plot the relevant portion of the $t = 12$ panel of fig.
\ref{defvac}, where these features are clearly distinguishable.

\begin{figure}[th!]
  \begin{center}
    \includegraphics[width=0.7\textwidth]{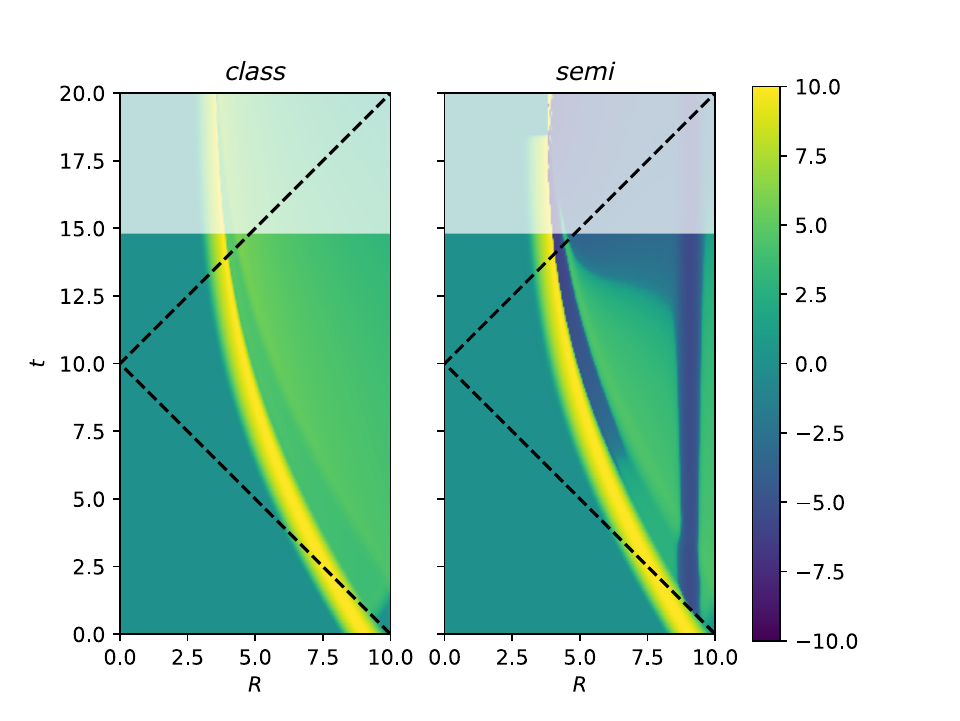}
\end{center}
\caption{\label{evolvedef}Contour plot the function $\tmop{sign}
  (\mathcal{h}) \ln (1 + 10^4 | \mathcal{h} |)$ of the Hamiltonian density
  $\mathcal{h}$ for the time evolution of our default system. The shaded area
  on top corresponds to the unsafe zone $t \gtrsim 14.8$. The black dashed
  diagonal lines correspond to $\mathrm{d}r = \pm \mathrm{d}t$ and indicate
  the trajectory of a radially infalling massless test particle on a fictitious
  flat background metric.}
\end{figure}

In order to give a visual illustration of the time evolution of our default
system, we provide in fig.~\ref{evolvedef} a contour plot the hamiltonian
density $\mathcal{h}$ for te time evolution up to $t = 20$ that compares the
classical to the semiclassical evolution.

\begin{figure}[th!]
  \begin{center}
    \includegraphics[width=0.7\textwidth]{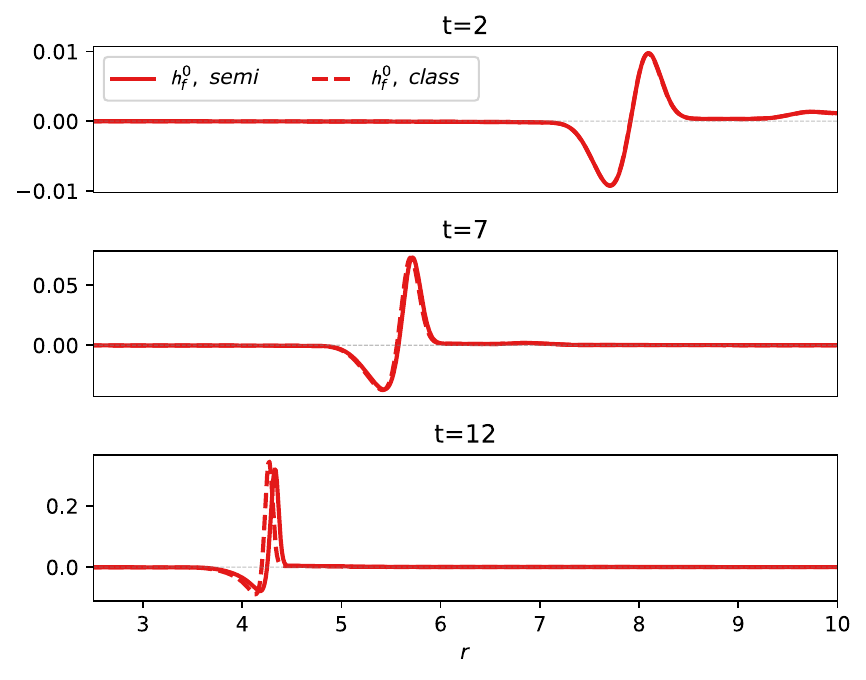}
    \includegraphics[width=0.7\textwidth]{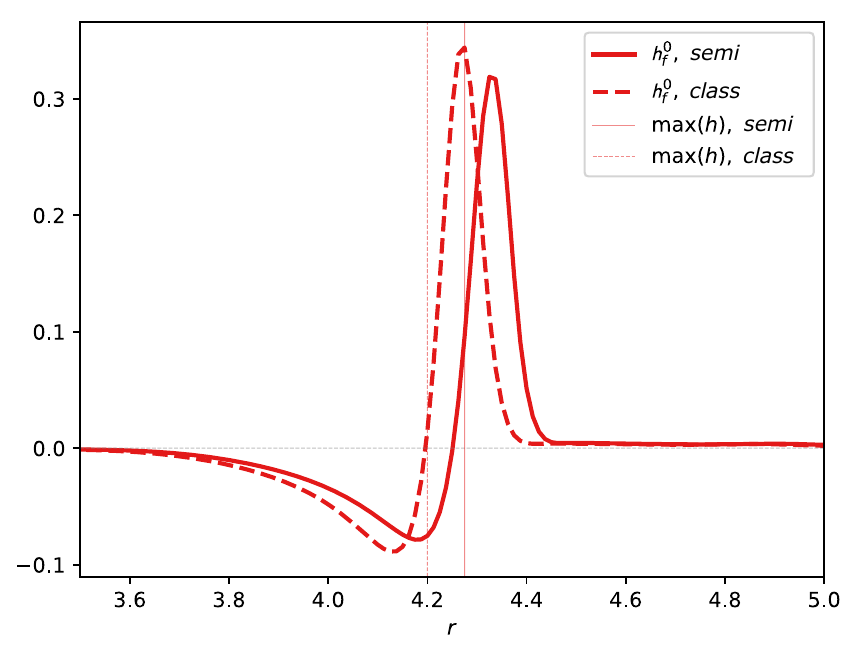}
\end{center}
\caption{\label{finh0}Time evolution of the final state vacuum
  contribution to the hamiltonian density
  $\mathcal{h}^0_f$ for three different times (top) and a detail from
  $t=12$ (bottom). Horizontal lines indicate the peak position of
  the total hamiltonian density $\mathcal{h}$. Form the perspective ot
  the final state vacuum, quantum effects deplete the energy density
  inside the forming horizon and enhance it outside.}
\end{figure}

\begin{figure}[th!]
  \begin{center}
    \includegraphics[width=0.7\textwidth]{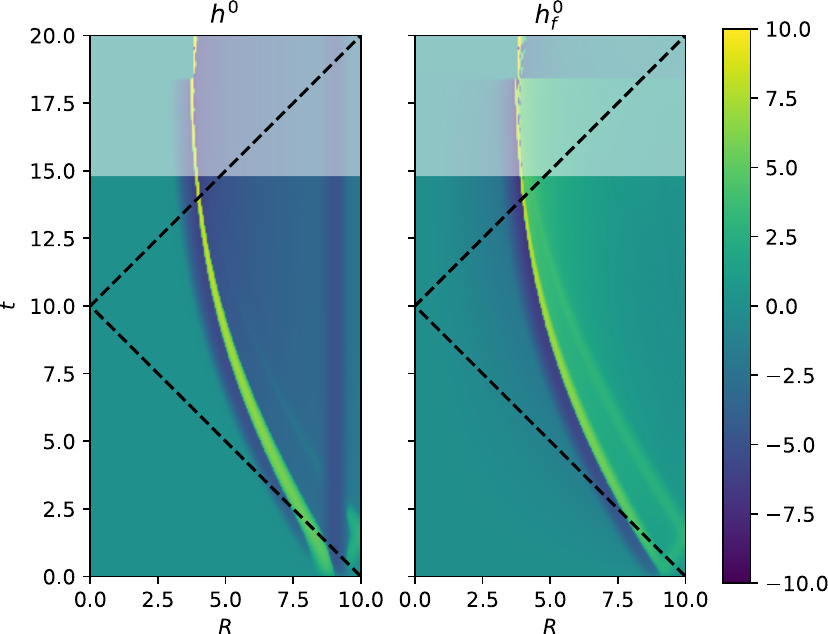}
\end{center}
\caption{\label{comph0} Comparison of $\mathcal{h}^0$ with
  $\mathcal{h}^0_f$. In each case we plot the function $\tmop{sign}
  (x) \ln (1 + 10^4 |x |)$ where $x$ is either $\mathcal{h}^0$ or
  $\mathcal{h}^0_f$. The shaded area
  on top corresponds to the unsafe zone $t \gtrsim 14.8$. The black dashed
  diagonal lines correspond to $\mathrm{d}r = \pm \mathrm{d}t$ and indicate
  the trajectory of a radially infalling massless test particle on a fictitious
  flat background metric.}
\end{figure}

The final observable we would like to present is the ``final state''
vacuum contribution to the hamiltonian density $\mathcal{h}^0_f$. We
define it as the vacuum contribution to the hamiltonian density, but
with a vacuum subtraction term (\ref{nof}) related to the vacuum
that corresponds to the metric parameters
$\hat{\alpha}=\hat{\alpha}(t)$ and $d=d(t)$ at the given time $t$ in
the evolution. We plot this quantity in fig.~\ref{finh0} for three
different times. As can be
seen, according to the final state vacuum, quantum effects lead to a
depletion of energy inside the forming horizon and an enhancement
outside.  In fig.~\ref{comph0} we compare the vacuum contribution to the
hamiltonian density $\mathcal{h}^0$ with its final state counterpart $\mathcal{h}^0_f$.

\subsection{Radial discretization}\label{contlim}

\begin{figure}[th!]
  \begin{center}
    \includegraphics[width=0.7\textwidth]{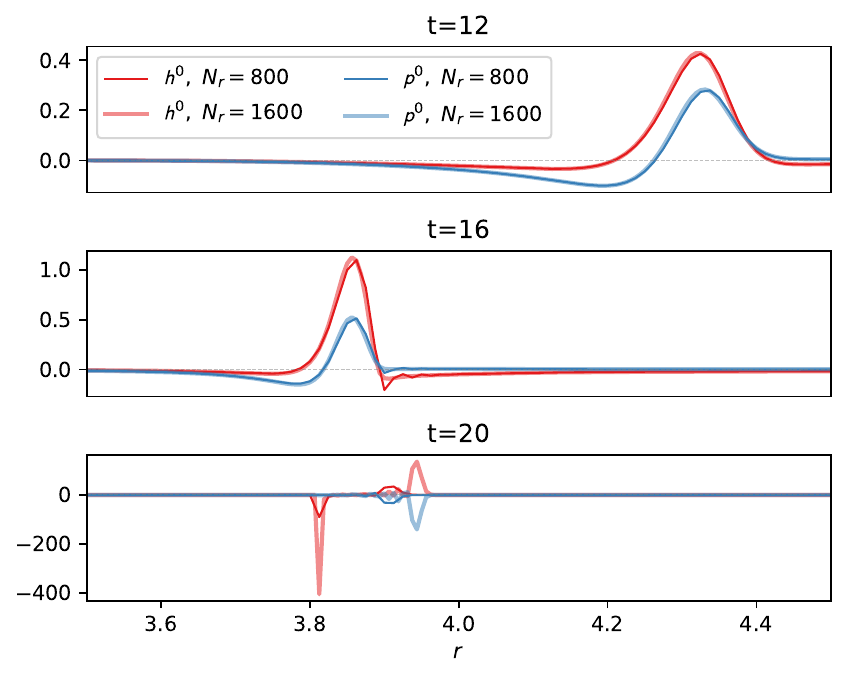}
  \end{center}
  \caption{\label{outofsafe}Comparison between two different radial
  discretizations $N_r = 800, 1600$ of the same physical system at different
  times $t$. The top panel represents a time that is in the safe zone of both
  discretizations, while the middle panel corresponds to a time that is out of
  the safe zone of the coarser $N_r = 800$ discretization while still being
  inside the safe zone of the finer $N_r = 1600$ case. The bottom panel shows
  a time that is out of the safe zone even for the finer discretization. }
\end{figure}

If we look at the time evolution of the default case beyond the safe zone $t
\gtrsim t_{\tmop{safe}}$, we see the gradual onset of radial discretization
artefacts, mainly through high frequency modes. This behaviour is displayed in
fig.~\ref{outofsafe}, where we compare the results for our default case to
those of the same system but with a radial discretization $\Delta$ that is
twice as fine, i.e. $N_r = 1600$.

\begin{figure}[th!]
  \begin{center}
    \includegraphics[width=0.7\textwidth]{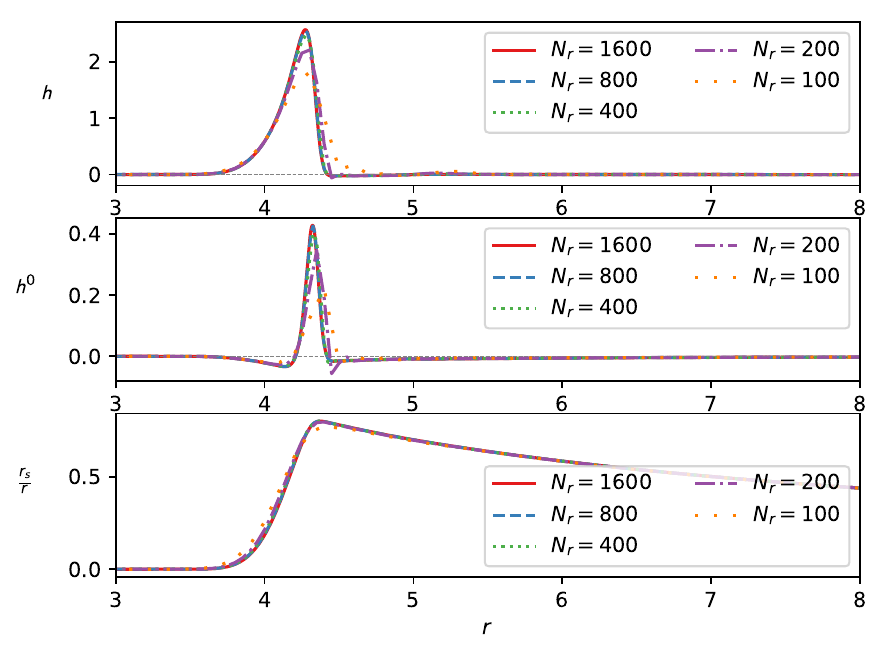}
\end{center}
\caption{\label{contlimfig}Comparison of different radial discretizations at
  $t = 12$. Oe can see that for $N_r = 100, 200$ this time is outside the safe
  zone while it is still inside for the finer discretizations.}
\end{figure}

For this finer discretization the safe zone extends to $t \lesssim 17.5$. We
can clearly see that when we are within the safe zone of both systems, the
results of both discretizations agree (top panel of fig.~\ref{outofsafe}). At
a time that is out of the safe zone for the coarser discretization but still
within the safe zone of the finer one (middle panel), the coarser
discretization starts to exhibit high frequency discretization artefacts while
the finer discretization does not. Finally, when we are out of the safe zone
of both systems (bottom panel), the finer discretization shows high frequency
discretization artefacts as well. We have checked that this behaviour is
generically true also for the even coarser discretizations, which is why we
believe (\ref{safezone}) to be a reasonable criterion for the specific bump
shape we use. In fig.~\ref{contlimfig} we plot the Hamiltonian density
$\mathcal{h}$, its vacuum contribution $\mathcal{h}^0$ and $r_s / r$ for all
five radial discretizations at the time $t = 12$. At this time we are,
according to tab. \ref{tsafe}, out of the safe zone for our two coarsest
discretizations with $N_r = 100, 200$ but within the safe zone of all
discretizations with $N_r = 400$ or finer.

\begin{figure}[th!]
  \begin{center}
    \includegraphics[width=0.7\textwidth]{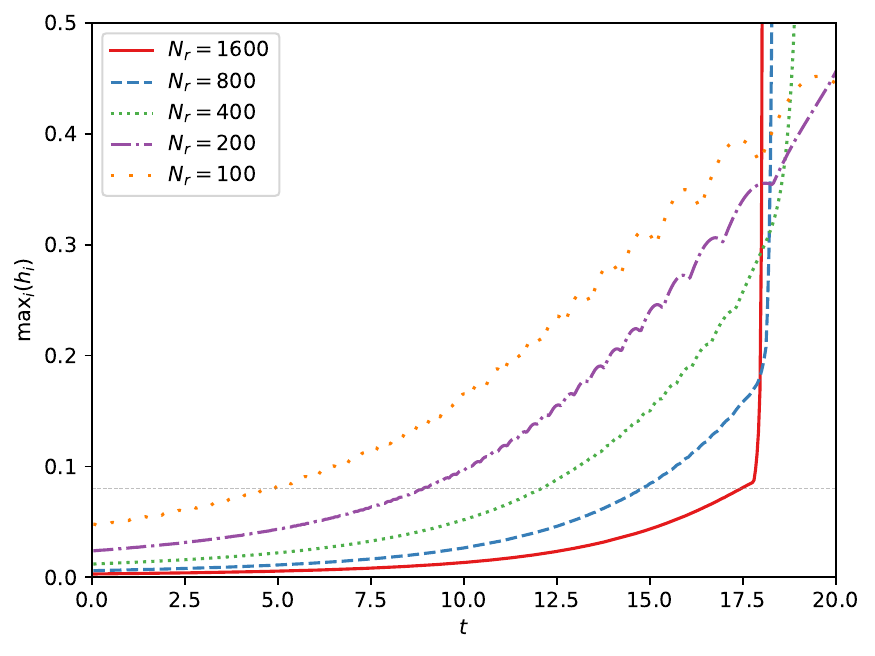}
\end{center}
\caption{\label{szplot}Comparison of the algorithmically relevant quantity
  $\max_i (h_i)$ for different radial discretizations. The horizontal line is
  at $0.08$, corresponding to the criterion (\ref{safezone}).}
\end{figure}

Finally, we turn again to our safe zone criterion (\ref{safezone}). In fig.
\ref{szplot} we show the time evolution of $\max_i (h_i)$ for the various
discretizations from which we can read off the extent of the safe zone
$t_{\tmop{safe}}$ that we give in tab. \ref{tsafe}.

Interestingly, there is a marked increase of $\max_i (h_i)$ just barely
outside the safe zone of our finest discretization. If this would happen in a
region where we could trust our simulation, it would be indicative of a
forming horizon. Since it is beyond the safe region however, we can not
exclude that it is purely a discretization artefact. Nonetheless, in
the semiclassical
simulation a horizon is actually forming around this time and we see $d
\rightarrow 0$ there.

\begin{figure}[th!]
  \begin{center}
    \includegraphics[width=0.7\textwidth]{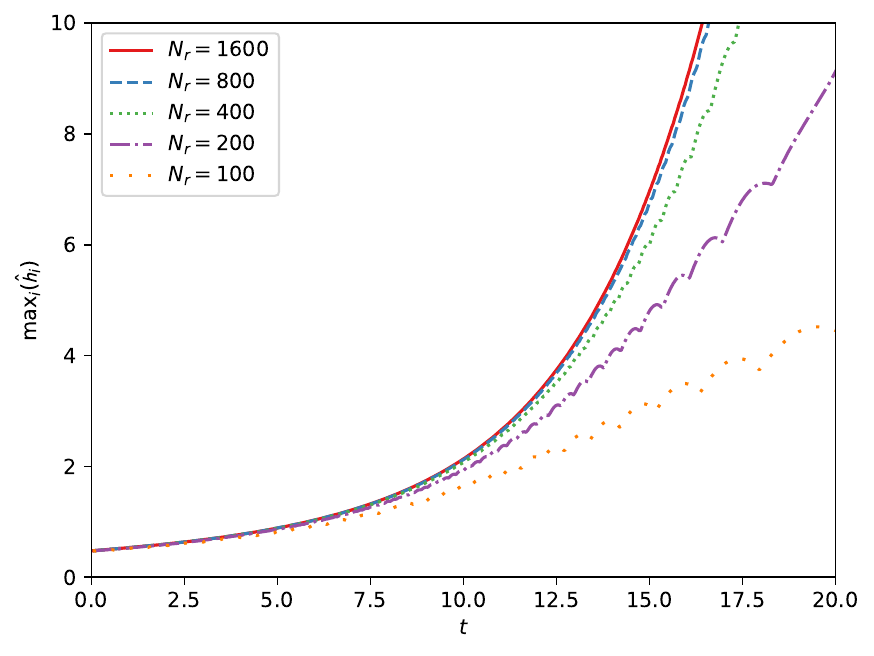}
\end{center}
\caption{\label{comphhat}Comparison of the physically relevant quantity
  $\max_i (\hat{\mathcal{h}}_i)$ for different radial discretizations.}
\end{figure}

The $\max_i (h_i)$ that we have plotted in fig.~\ref{szplot} is relevant for
algorithmic consideration, but not directly suitable for physical comparisons
as it contains the radial discretization length $\Delta$ via $h_i =
\hat{\mathcal{h}}_i \Delta$. We can instead compare $\max_i
(\hat{\mathcal{h}}_i) = \max_i (h_i) / \Delta$ that is a physical density.
This comparison is depicted in fig.~\ref{comphhat}. We see that initially all
curves agree and that the coarser discretizations branch off from the rest of
the curves around their respective $t_{\tmop{safe}}$. Interestingly, it seems
that outside of the range of validity, the coarser the discretization the less
likely it is to lead to horizon formation.

\

\subsection{Discretizing the derivative operator}

In the default case we use the simple forward difference operator
\[ (\nabla_f)_{i j} = \frac{\delta_{i + 1, j} - \delta_{i, j}}{\Delta} \]
to discretize the derivative in (\ref{defq0}, \ref{qbar}). To estimate the
effect that this choice has on out result, we investigate three other choices.
The first two are the backward difference operator $\nabla_b = \nabla_f^T$ and
the symmetric difference operator $\nabla = (\nabla_f + \nabla_b) / 2$. Both
the forward and the backward difference operator have discretization effects
that are formally of the order $\partial - \nabla_{f / b} = O (\Delta)$ while
for the symmetric difference operator we have \ $\partial - \nabla = O
(\Delta^2) .$ In addition to these two choices, we also investigate a more
extended difference operator which involves two points in forward and one in
backward direction
\[ (\nabla_x)_{i j} = \frac{1}{\Delta} \sum_{k = - 1}^2 c_k \delta_{i + x, j}
\]
where the $c_k$ are fixed such that the discretization effects are $\partial -
\nabla_x = O (\Delta^3)$.

\begin{figure}[th!]
  \begin{center}
    \includegraphics[width=0.7\textwidth]{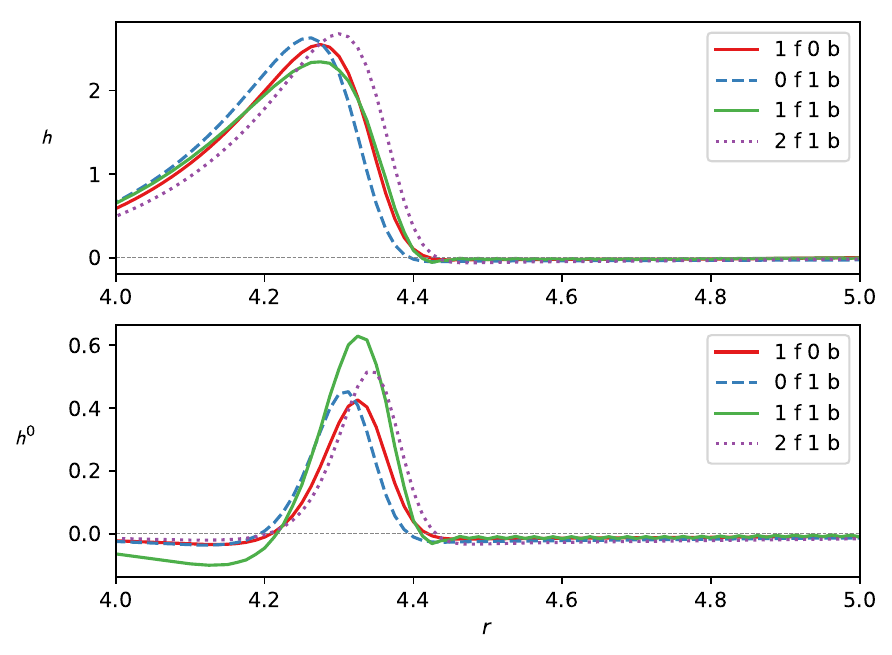}
  \end{center}
  \caption{\label{discderpl}Hamiltonian density \ $\mathcal{h}$ and its vacuum
  contribution $\mathcal{h}^0$ at time $t = 12$ for the semiclassical case and
  various discretizations of the derivative operator. The labels correspond to
  the number of forward (f) and backward (b) points included in the
  discretized derivative.}
\end{figure}

\begin{figure}[th!]
  \begin{center}
    \includegraphics[width=0.7\textwidth]{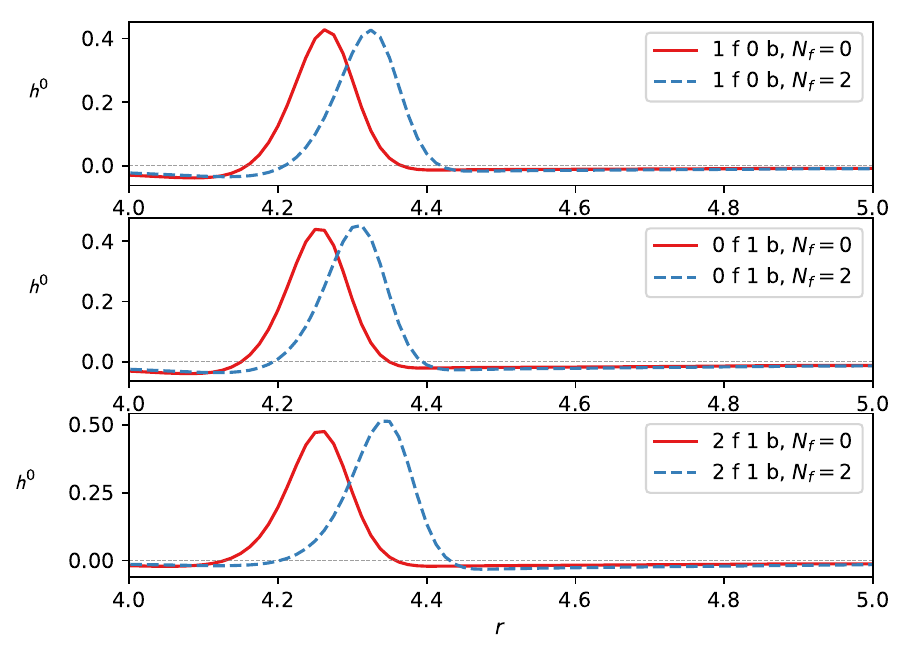}
\end{center}
\caption{\label{disc02}Comparison of the vacuum contribution to the
  Hamiltonian density $\mathcal{h}^0$ between the classical and semiclassical
  cases for three different discretizations of the derivative operator. Note
  that the important qualitative features, i.e. the different peak positions
  for the classical and semiclassical cases and the enhancement of
  $\mathcal{h}$ by vacuum contributions in the peak region is clearly present
  for all choices of the discretization.}
\end{figure}

In fig.~\ref{discderpl} we compare the Hamiltonian density $\mathcal{h}$ and
its vacuum contribution $\mathcal{h}^0$ at time $t = 12$ for all
discretizations. One can see that despite the differences the qualitative
behaviour of all four discretizations is the same. We also note that for the
symmetric difference operator we see a marked increase of high frequency
discretization artefacts. This is not surprising, as the symmetric difference
operator has problems disentangling low and very high frequency
modes.\footnote{We can see a related phenomenon in the simple case of applying
the symmetric difference operator to a Fourier mode $f_k (x) = \exp (- i k x)$
which results in $(\nabla f) (x) = i \frac{\sin (\Delta k)}{\Delta} f (x)$.
Thus to the operator $\nabla$ the Fourier modes of momentum $k$ and $\pi /
\Delta - k$ are indistinguishable. More concretely, the exclusive
even-odd coupling of the operator $q^0$ enforces opposing checkerboard
patterns of the coefficient matrices $u$ and $v$. Among other things,
this forces the diagonal elements of $u^{\dag}v$ to vanish, which in
the light of the continuity equation (\ref{conteq}) is a particularly
unwelcome discretization artefact.} For the remaining three discretizations we
show in fig.~\ref{disc02} the difference between the classical and
semiclassical $\mathcal{h}^0$ at $t = 12$. We can clearly see that the
feature of a more outward lying semiclassical peak that is enhanced by quantum
effects is equally present for all three discretizations of the derivative
operator.

\begin{figure}[th!]
  \begin{center}
    \includegraphics[width=0.7\textwidth]{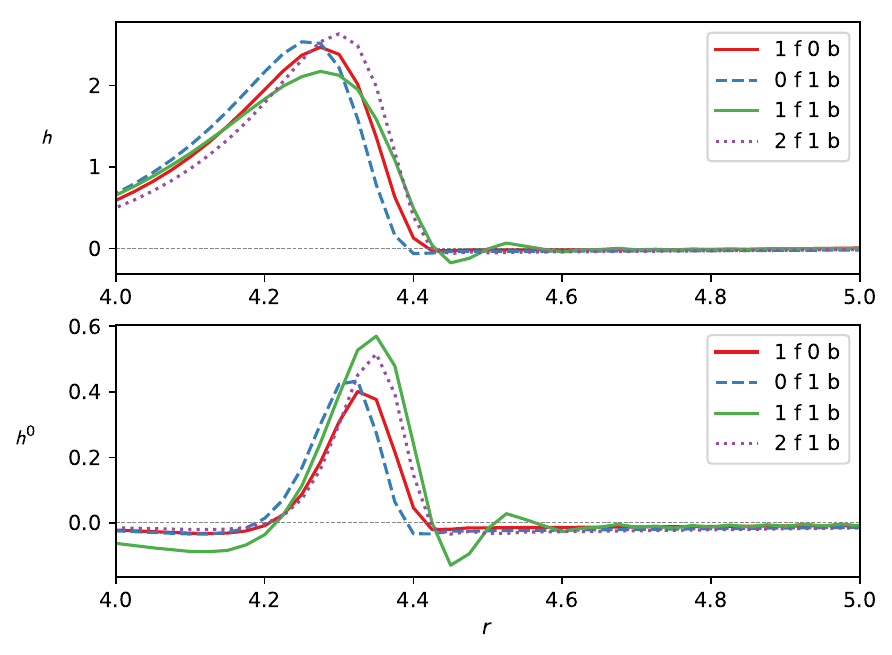}
\end{center}

\begin{center}
  \includegraphics[width=0.7\textwidth]{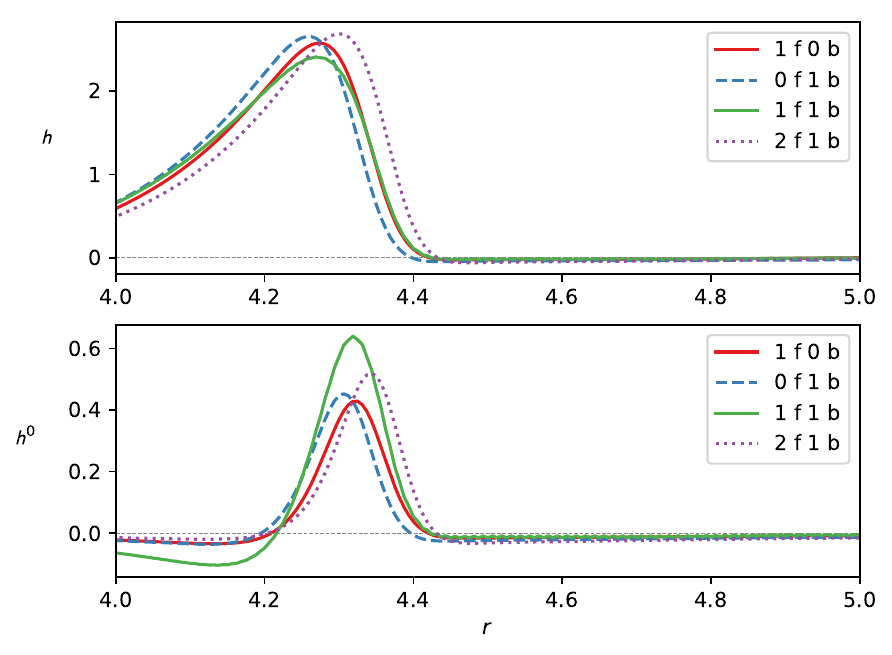}
\end{center}
\caption{\label{discdisc}Same as fig.~\ref{discderpl} for two different
  radial discretizations corresponding to $N_r = 400$ (top) and $N_r =
  1600$(bottom).}
\end{figure}

The difference between the discretizations is of course a discretization error
that should vanish in the limit $\Delta \rightarrow 0$. In fig.~\ref{discdisc}
we therefore show the comparison of fig.~\ref{discderpl} again, but this time
for one coarser $N_r = 400$ and one finer $N_r = 1600$ radial discretization.
We can see that the agreement between the different discretizations of the
derivative operator is slightly better for the finer radial discretization.
Nonetheless a more detailed study of this topic on substantially finer radial
discretizations would be highly desirable.

\subsection{Other variations of the update procedure}

\begin{figure}[th!]
  \begin{center}
    \includegraphics[width=0.7\textwidth]{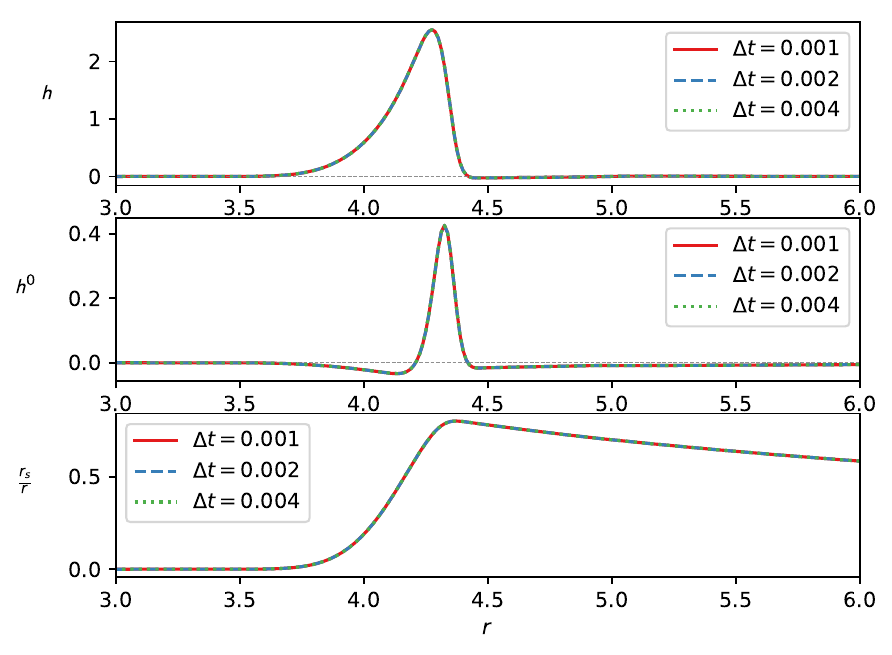}
\end{center}
\caption{\label{tempdisc}Comparison of results for different time steps at
  $t = 12$.}
\end{figure}

In addition to the radial discretization, we also have to check that the
temporal discretization we use is fine enough. In fig.~\ref{tempdisc} we
compare results for three different time steps and find that they are
virtually indistinguishable. For our finest radial discretization $N_r = 1600$
we actually observe some tiny discrepancies in the high frequency components
around the position of the original peak, but even they are entirely
negligible for the entire time evolution even beyond the safe zone.

\begin{figure}[th!]
  \begin{center}
    \includegraphics[width=0.7\textwidth]{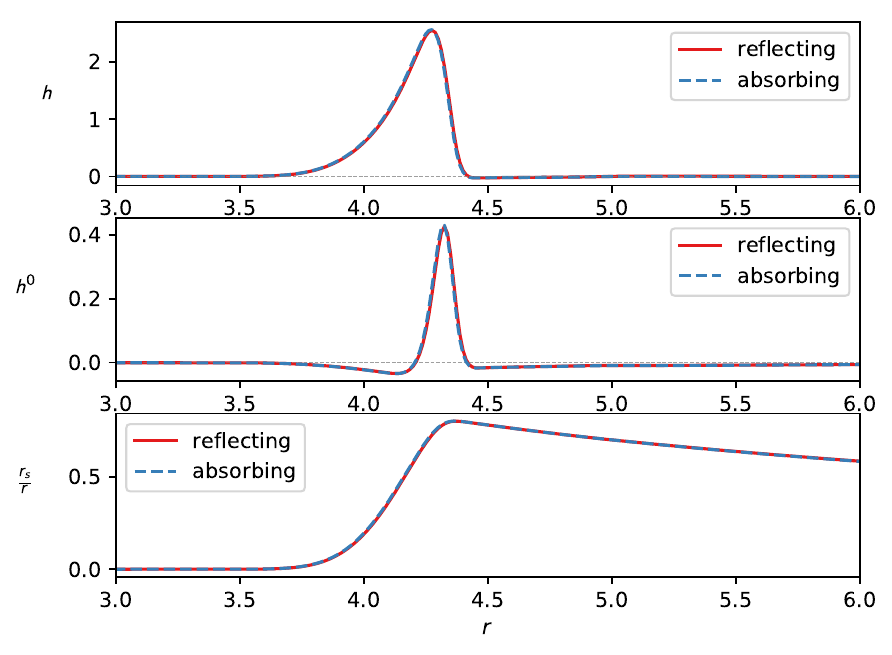}
  \end{center}
  \caption{\label{abref10}Comparison of results with different boundary
  conditions at $t = 12$.}
\end{figure}

Next we would like to investigate the effect of classically absorbing boundary
conditions as introduced in section \ref{absorbing_bc}. In fig.~\ref{abref10} we compare the time evolution between the
two set of boundary conditions and see that they agree rather precisely.

\begin{figure}[th!]
  \begin{center}
    \includegraphics[width=0.7\textwidth]{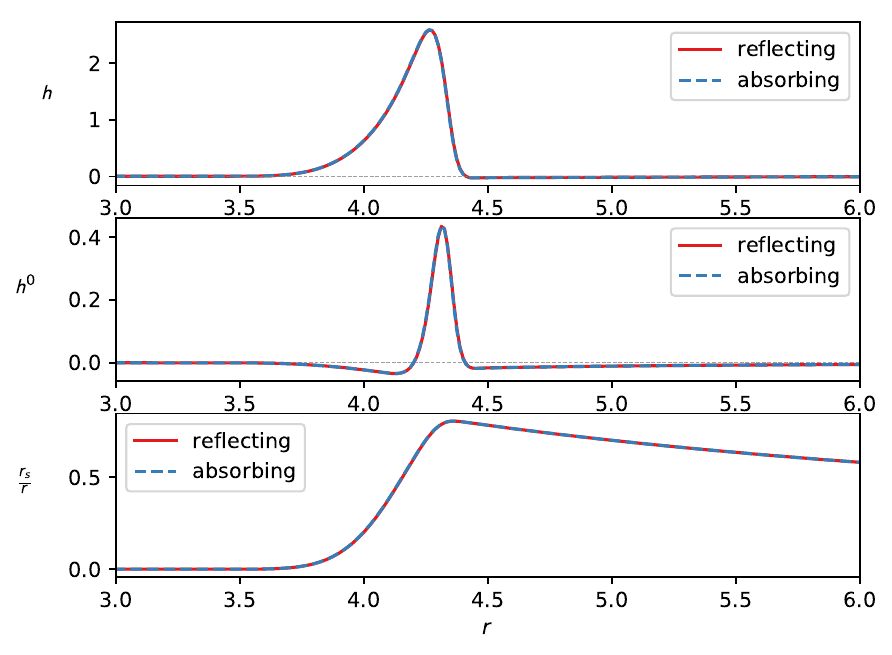}
  \end{center}
  \caption{\label{abref12}The same plot as fig.~\ref{abref10} but for a larger
  discretized volume $r_{N_r} = 12$.}
\end{figure}

The slight difference between the results with the two boundary conditions in
fig.~\ref{abref10} should of course be a finite volume effect. We can check
this explicitly by performing the same comparison on a larger discretized
volume with $r_{N_r} = 12$. We do this in fig.~\ref{abref12} and indeed see
that the difference between the two cases gets even smaller.

\begin{figure}[th!]
  \begin{center}
    \includegraphics[width=0.7\textwidth]{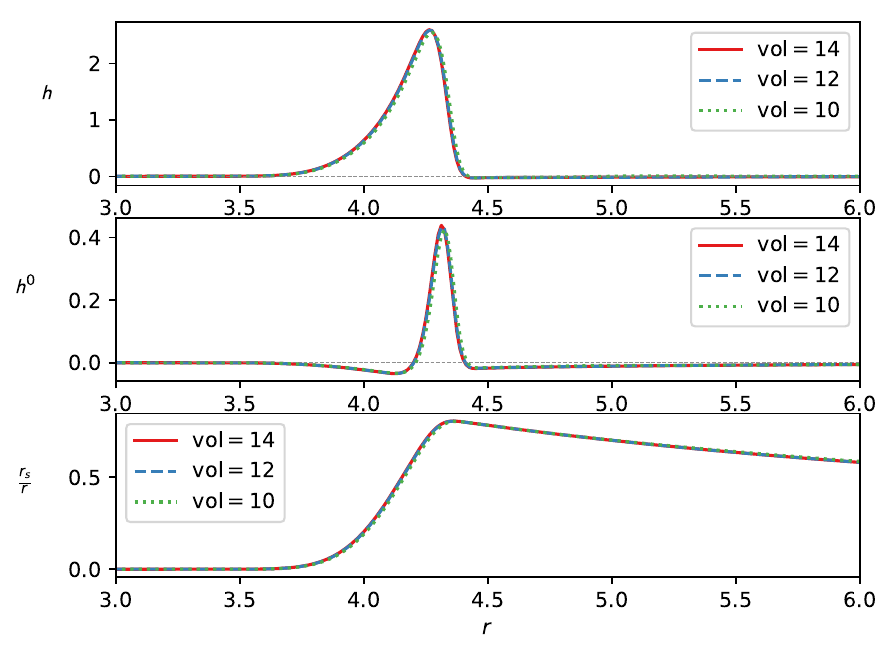}
  \end{center}
  \caption{\label{fvr}Comparison of our default system of radial extent
  $r_{N_r} = 10$ at $t = 12$ with two systems of larger radial extent $r_{N_r}
  = 12, 14$ that are otherwise identical.}
\end{figure}

We can also study finite volume effects by comparing our default system with
two systems that have a larger $r_{N_r} = 12$ and 14. As seen in fig.
\ref{fvr}, finite volume effects are very small.

\begin{figure}[th!]
  \begin{center}
    \includegraphics[width=0.7\textwidth]{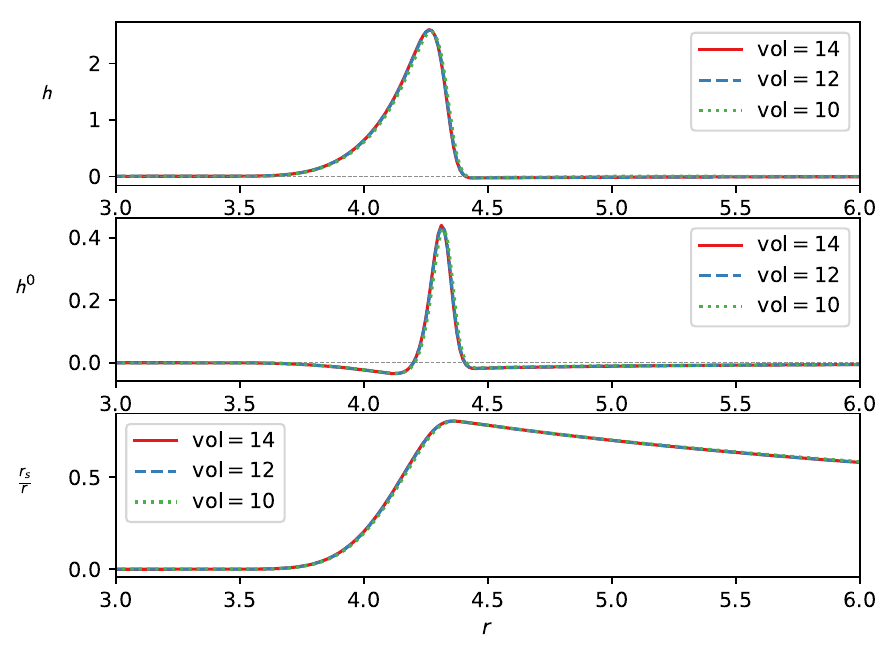}
  \end{center}
  \caption{\label{fva}The same as fig.~\ref{fvr} but with absorbing boundary
  conditions.}
\end{figure}

In fig.~\ref{fva} we repeat this comparison for absorbing boundary conditions.
We see that finite volume effects are even smaller in this case, indicating
that the main contribution towards finite volume effects for our default
system in fact comes from classical reflection off the outer boundary.

\begin{figure}[th!]
  \begin{center}
    \includegraphics[width=0.7\textwidth]{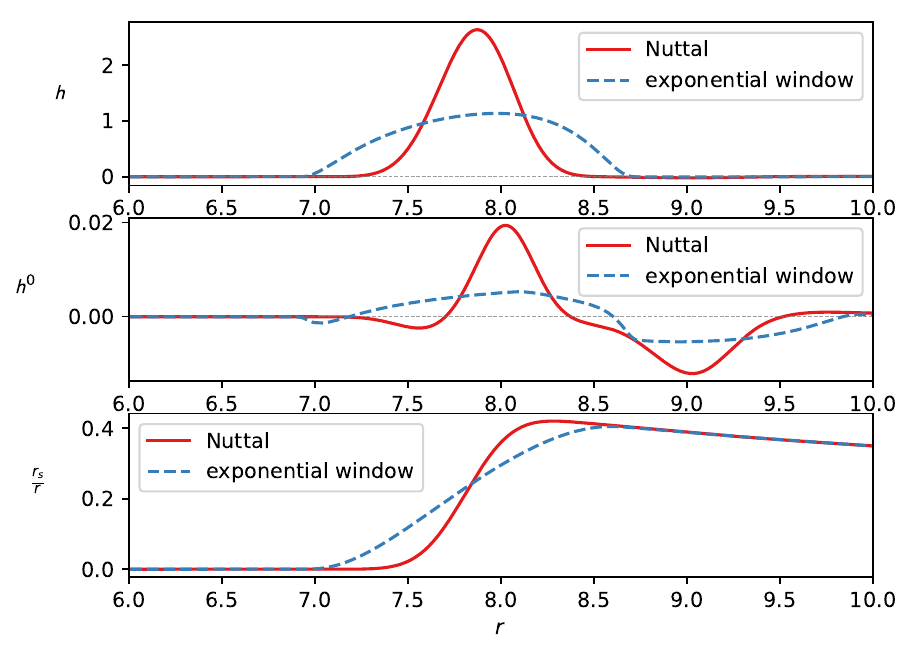}
\end{center}
\caption{\label{bmp1}Comparison the semiclassical evolution with different
  initial scalar field bump shapes at $t = 2$.}
\end{figure}

Next we investigate varying the shape of the initial $\hat{\mathcal{h}}$ of the
scalar field from the usual Nuttall form (\ref{nut}) to the exponential form
(\ref{bmp}). Fig.~\ref{bmp1} compares the two systems at an early time $t =
2$. Clearly the exponential shape is less peaked than the Nuttall form, but
the important qualitative feature, the enhancement of the Hamiltonian density
$\mathcal{h}$ around the peak region by the vacuum contribution
$\mathcal{h}^0$, is also prominent for the exponential bump.

\begin{figure}[th!]
  \begin{center}
    \includegraphics[width=0.7\textwidth]{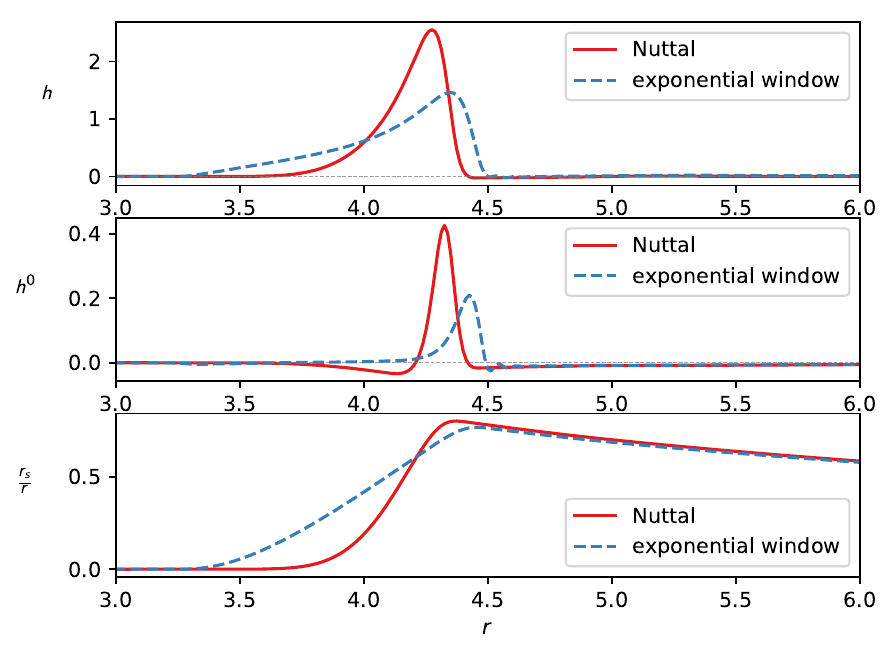}
\end{center}
\caption{\label{bmp2}The same as fig.~\ref{bmp1} but at a later time $t =
  12$.}
\end{figure}

At a later time $t = 12$ this property persists as is evident from fig.
\ref{bmp2}. For the exponential bump $\mathcal{h}$ peaks at a larger radius,
but the vacuum contribution $\mathcal{h}^0$ does enhance the peak as it does
for the Nuttall bump.

\begin{figure}[th!]
  \begin{center}
    \includegraphics[width=0.7\textwidth]{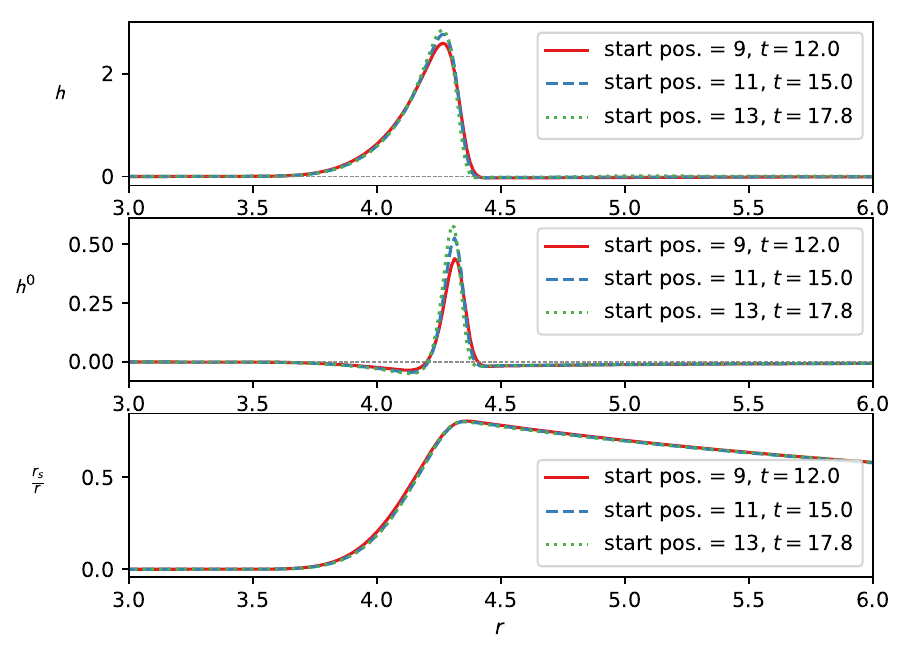}
  \end{center}
  \caption{\label{bumpout}Comparison of different starting positions of the
  scalar field. Results are compared at times where the respective systems
  have reached the same peak position of $\mathcal{h}$ than the reference $t =
  12$ for our default system. Fields that started further out show a more
  pronounced peak of both $\mathcal{h}$ and its vacuum contribution
  $\mathcal{h}^0$.}
\end{figure}

The next topic we would like to address is that of the error made by putting
the scalar field at a rather small distance initially instead of
asymptotically far away. We address this issue in a direct way by increasing
the volume and putting the center of the initial bump to a larger distance
and, more indirectly, by varying the vacuum subtraction. The effect of moving
the initial scalar field further outward on a larger system is displayed in
fig.~\ref{bumpout}. Since the shape of the bump changes as it propagates
towards the center, we do not expect the result from the different starting
positions to agree exactly at the respective times when they have reached the
same radial coordinate. Nonetheless we can see that the differences are
minimal with the bumps that started from a larger distance showing higher
peaks.

\

\begin{figure}[th!]
  \begin{center}
    \includegraphics[width=0.7\textwidth]{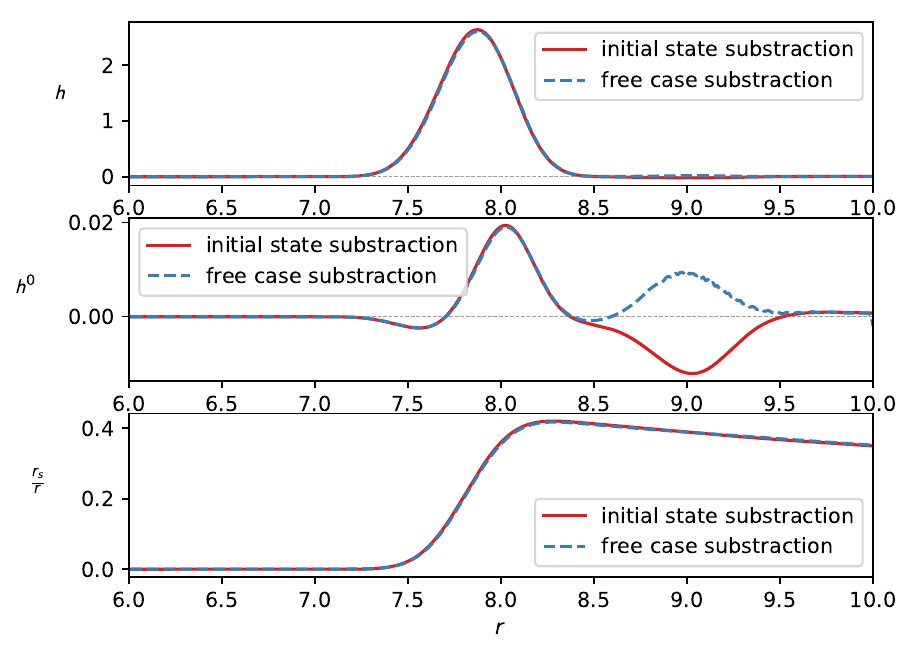}
\end{center}
\caption{\label{vst2}Comparison of two vacuum subtraction procedures at time
  $t = 2$. The pronounced difference in $\mathcal{h}^0$ between the two
  procedures is confined to the region of the initial bump position.}
\end{figure}

The effect of changing the vacuum subtraction from the standard initial state
normal ordering (\ref{no1}) to the free field subtraction (\ref{no2}) modified
as described around (\ref{vss}) at an early time $t = 2$ is shown in fig.
\ref{vst2}.

\begin{figure}[th!]
  \begin{center}
    \includegraphics[width=0.7\textwidth]{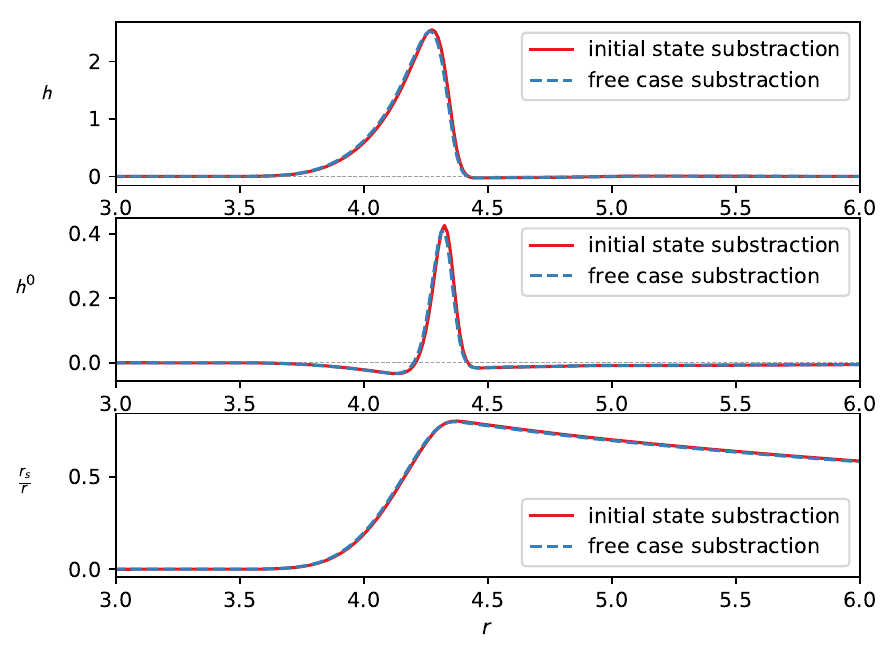}
\end{center}
\caption{\label{vst12}The same as fig.~\ref{vst2}, but at a later time $t =
  12$.}
\end{figure}

Although we can see a marked difference in the vacuum contribution to the
Hamiltonian density $\mathcal{h}^0$ around the initial position of the scalar
field, the effect on $\mathcal{h}$ and even on $\mathcal{h}^0$ outside this
region is negligible. The difference between the two procedures stays small
also at later times, as seen in fig.~\ref{vst12} for $t = 12.$

\begin{figure}[th!]
  \begin{center}
    \includegraphics[width=0.7\textwidth]{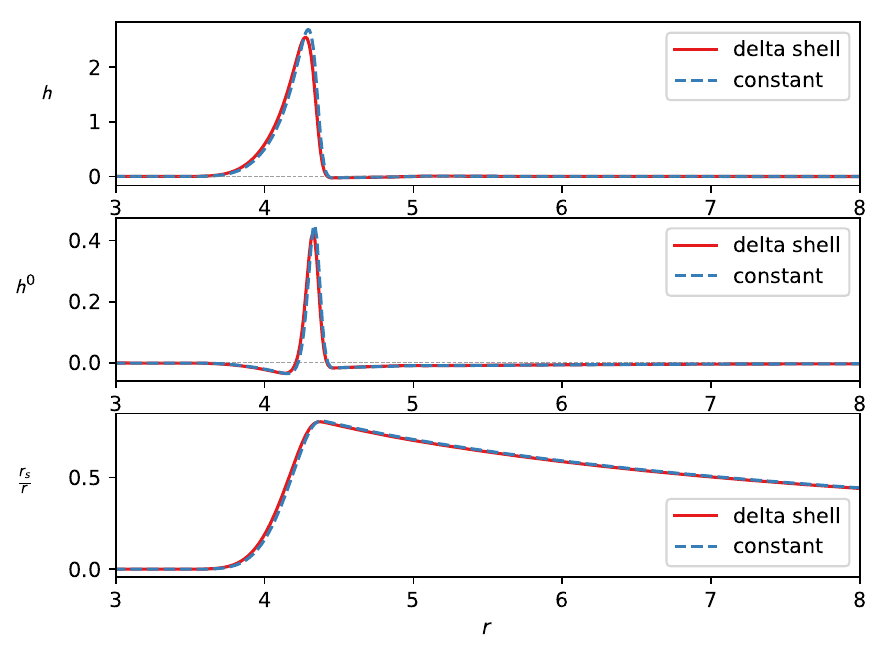}
\end{center}
\caption{\label{intcomp}Comparison of the time evolution with our two radial
  integration schemes at $t = 12$.}
\end{figure}

Finally, we would like to investigate the effect of changing the radial
integration scheme from the standard $\delta$-shell case (\ref{intd1}) to the
piecewise constant (\ref{intd2}). As can be seen in fig.~\ref{intcomp} also
this effect is entirely negligible at $t = 12$.

\subsection{The operators $q^0$ and $\overline{q}$}

\begin{figure}[p]
  \begin{center}
    \includegraphics{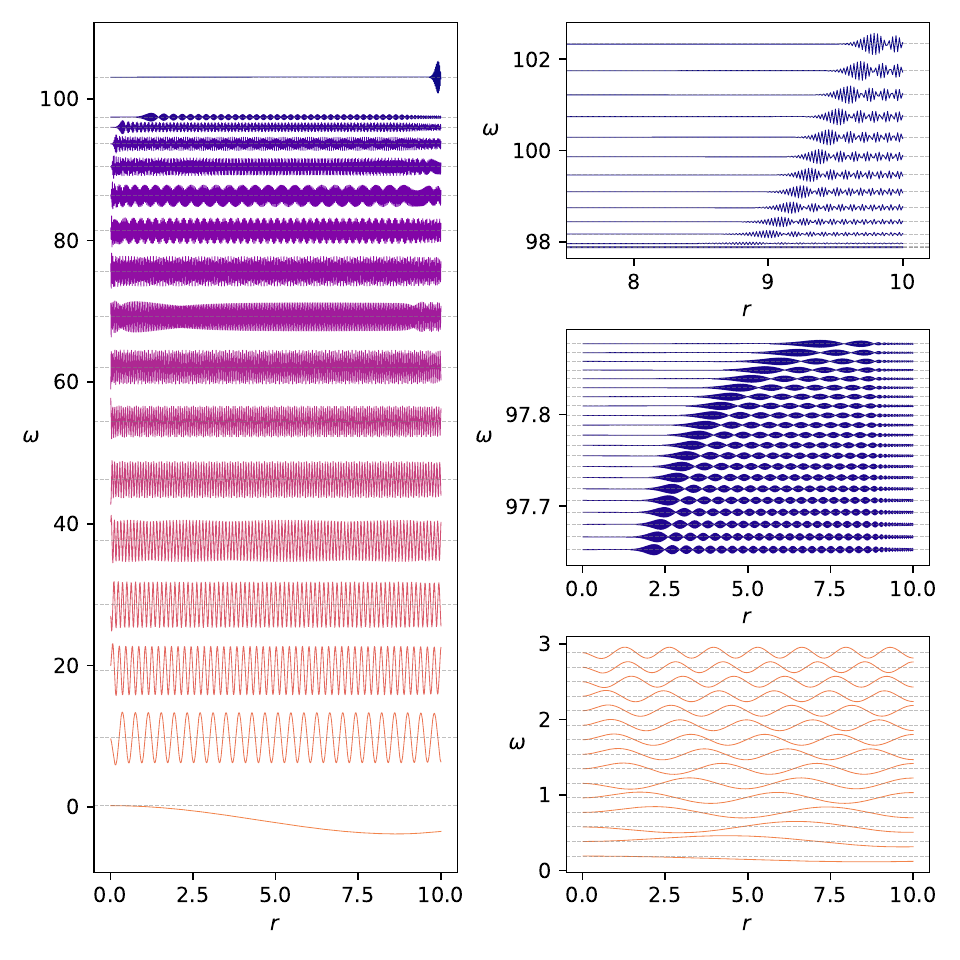}
  \end{center}
  \caption{\label{vmodes}Some of the singular vectors $U_{\cdummy k}$ of the
  operator $q^0$ for our default system and their singular values $\omega_k$.
  On the left hand side we plot every $50^{\tmop{th}}$ mode starting from the
  mode with the lowest singular value and the mode with the highest singular
  value. On the right hand side we zoom in on three interesting regions for
  which we plot every mode. Note that the spacing of the modes decreases for
  increasing $\omega$ except for the few modes with the highest $\omega$ that
  are concentrated in the region of the original bump position.}
\end{figure}

It is quite instructive to investigate a bit the singular mode structure of
the operators $q^0$ (\ref{defq0}) and $\overline{q}$ (\ref{qbar}). We start
with the singular vectors $U_{\cdummy k}$ and $V_{\cdummy k}$ of $q^0$, which
are interesting because they form the basis of the mode expansion. We plot a
sample of the $V_{\cdummy k}$ in fig.~\ref{vmodes}. It is interesting that the
shape of the initial bump is reflected in the modes with largest $\omega$,
since they have support almost entirely in that region.

\begin{figure}[th!]
  \begin{center}
    \includegraphics[width=0.7\textwidth]{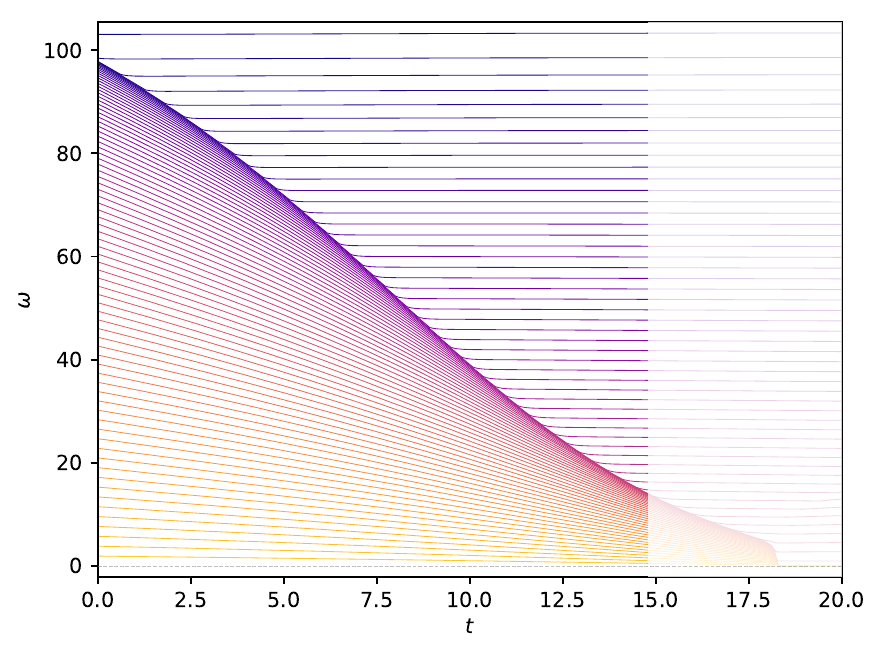}
\end{center}
\caption{\label{omevol}Time evolution of the singular values
  $\overline{\omega}_{}$ of $\overline{q}$. Starting from the highest, we plot
  every $10^{\tmop{th}}$ $\overline{\omega}$. The shaded area
  on the right marks the unsafe region $t > t_{\tmop{safe}}$.}
\end{figure}

Turning to the operator $\overline{q}$ that is used in the time evolution of
the scalar field, we are first interested in how its singular values change
during the course of the evolution of the system. This is plotted in fig.
\ref{omevol} for a representative selection of modes.

We see that during the course of the time evolution the number of low modes
increases while the high modes are thinned out. Deep within the unsafe region
we can even see a number of modes that reach singular values very close to
$\omega = 0$ in a very short time. It is reasonable to assume that this
concentration of low modes is connected to the onset of horizon formation. The
closer the system gets to forming a real horizon the more we expect modes to
separate into those that are largely outside the forming horizon and take part
in the usual time evolution and those modes that are largely inside and freeze
as seen in coordinate time $t$ (which is the physical time of an asymptotic
observer). In order check whether this picture is correct, we define the
``inside'' and ``outside'' components of the singular vectors belonging to the
singular value $\overline{\omega}_k$ as
\[ \begin{split}
     (f_V^{\tmop{in}})_k^2 = \sum_{i = 1}^s | \overline{V}_{s k} |^2 &  \qquad
     (f_V^{\tmop{out}})_k^2 = \sum_{i = s + 1}^{N_r} | \overline{V}_{s k}
     |^2\\
     (f_U^{\tmop{in}})_k^2 = \sum_{i = 1}^s | \overline{U}_{s k} |^2 &  \qquad
     (f_u^{\tmop{out}})_k^2 = \sum_{i = s + 1}^{N_r} | \overline{U}_{s k}
     |^2
   \end{split} \]
where we define $s$ to be the index of the radial coordinate with the minimum
$d_i / r_i$, i.e. the coordinate at which the system is closest to forming a
horizon. Since the singular vectors are unit normalized we have \
$(f_V^{\tmop{in}})_k^2 + (f_V^{\tmop{out}})^2_k = (f_U^{\tmop{in}})^2_k +
(f_U^{\tmop{out}})^2_k = 1$.

\begin{figure}[th!]
  \begin{center}
    \includegraphics[width=0.7\textwidth]{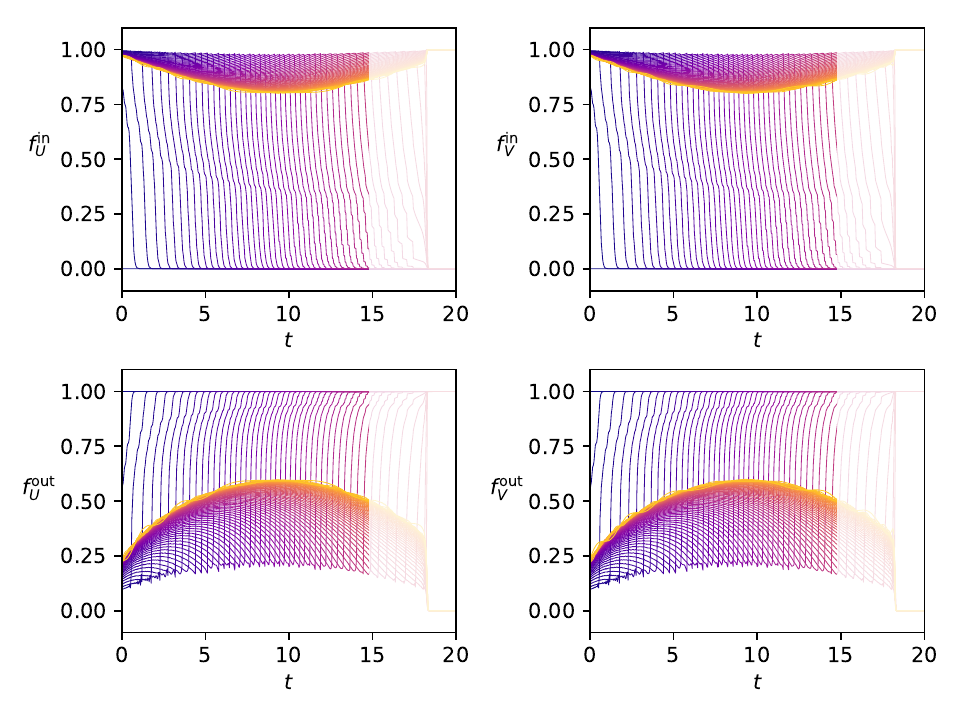}
  \end{center}
  \caption{\label{fio}The ``inside'' and ``outside'' components of the
  singular vectors of $\overline{q}$. We plot the same representative sample
  of motes as in fig.~\ref{omevol} with the same color coding.}
\end{figure}

In fig.~\ref{fio} we plot the $f_{U / V}^{\tmop{in} / \tmop{out}}$ of a
representative sample of modes. We see that with the exception of the few
modes with the very highest $\omega$, all modes start out with a substantial
``in'' component, which is of course what we expect from fig.~\ref{vmodes}. As
the bump and therefore the boundary between ``in'' and ``out'' moves towards
the center, we first see the bulk of the modes gaining a substantial ``out''
component. At around $t = 10$ however, we see that the ``out'' components of
the low modes start to decrease again while more and more of the higher modes
cross over to be fully in the ``out'' region. Finally, deep in the unsafe
region and at the point where fig.~\ref{omevol} showed the appearance of modes
with almost vanishing singular values, we see a sudden drop of the ``out''
components of these modes to almost zero, indicating that indeed they are
frozen behind the developing horizon.

In order to display the relevant properties of the eigenmodes more compactly
we define the mode separation parameter
\begin{equation}
  s^2 = \frac{1}{2} (\langle (f_U^{\tmop{in}})_k^2 + (f_V^{\tmop{in}})_k^2
  \rangle_{\tmop{in}} - \langle (f_U^{\tmop{in}})_k^2 + (f_V^{\tmop{in}})_k^2
  \rangle_{\tmop{out}}) \label{defms}
\end{equation}
where $\langle \cdummy \rangle_{\tmop{in}}$ denotes the average over all
``inside'' modes which we arbitrarily define as those modes with
$(f_U^{\tmop{in}})_k + (f_V^{\tmop{in}})_k > 1$. Correspondingly, $\langle
\cdummy \rangle_{\tmop{out}}$ denotes the average over all ``outside'' modes,
which are those that do not fulfill this condition. When a horizon forms, we
expect the the modes to divide into those that have support exclusively inside
the horizon and those that have support exclusively outside. Thus $s = 1$
indicates horizon formation.

\begin{figure}[th!]
  \begin{center}
    \includegraphics[width=0.7\textwidth]{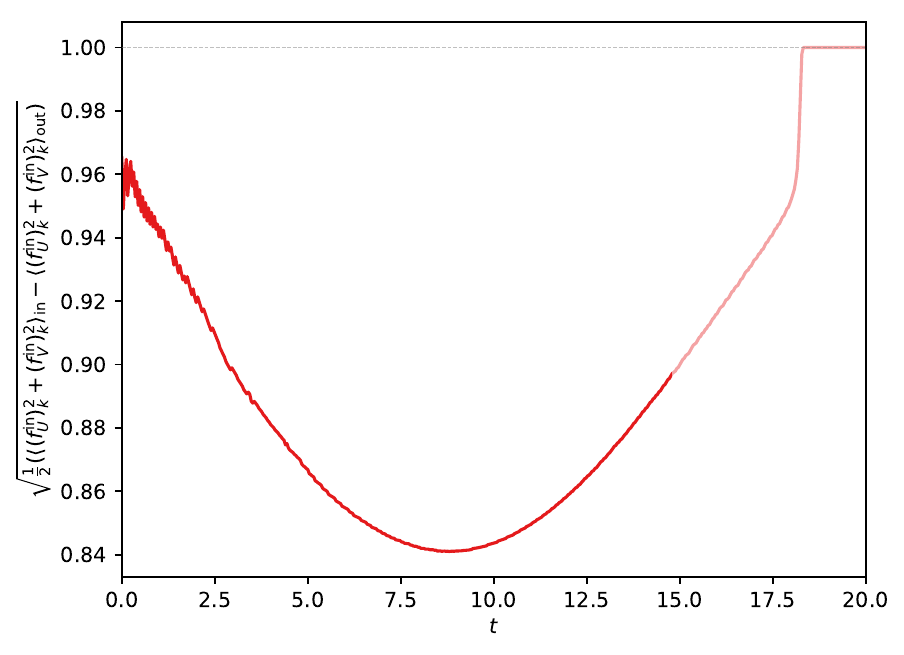}
\end{center}
\caption{\label{modesep}The mode separation parameter $s$ (\ref{defms}) vs.
  $t$. }
\end{figure}

In fig.~\ref{modesep} we plot the mode separation parameter vs. $t$. The
horizon formation is clearly visible in the unsafe zone.

\subsection{Consistency check}

In the continuum, the last equation in (\ref{eoms}), i.e.
\[ \dot{d} = - 2 \hat{\alpha} d \hat{\mathcal{p}}_r \]
together with the first one
\[ \frac{1 - d'}{d} = \frac{\hat{\alpha}'}{\hat{\alpha}} \]
guarantees that when $\hat{\mathcal{p}}_r$ vanishes in some finite region the
metric does not change there. Looking at our finite, discretized system as an
approximation to the infinite continuum case, we of course have
$\hat{\mathcal{p}}_r = 0$ in the region $r > r_{N_r}$, i.e. outside of our
outermost discrete coordinate. Consequently, if we had a perfect update
algorithm the metric parameters in the outside region and thus also $d_{N_r}$
would never change.\footnote{Remember that $\hat{\alpha}_p = 1$ is fixed by
our boundary condition for radial integration.} Since this property follows
from the one equation of motion that we never use explicitly, it is a useful
croscheck of the validity and accuracy of our numerical implementation.

\begin{figure}[th!]
  \begin{center}
    \includegraphics[width=0.7\textwidth]{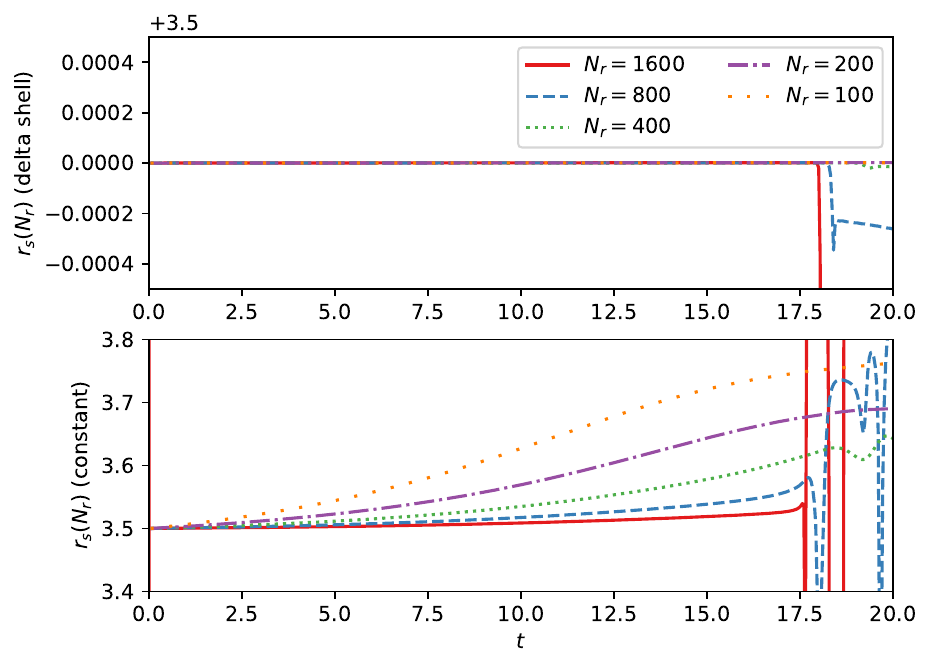}
\end{center}
\caption{\label{birk}Crosscheck of how constant the effective Schwarzschild
  radius $r_s = r_{N_r} - d_{N_r}$ is on the outermost shell.}
\end{figure}

In fig.~\ref{birk} we plot $r_s = r_{N_r} - d_{N_r}$ throughout the
semiclassical time evolution, varying the radial discretization and the radial
integration scheme. We see that $r_s$ is conserved to a relatively high
accuracy with the $\delta$-shell integration faring markedly better than the
piecewise constant. We can also see that for the finer discretizations the
agreement breaks down at large $t$ outside the safe region. This coincides
with the explosion of the maximum $h_i$ and the sudden separation of the
eigenmodes that we had observed above.

\

\subsection{Prospects of observing Hawking radiation}

An obvious question at this point is whether there is any prospect, in our
formalism, to observe the effects of Hawking radiation. Since we have no method
at present to represent the state of the scalar field in the outgoing Fock
space, the individual field quanta are of course inaccessible. We could
however hope to see the related outgoing energy flux. We can do a quick order
of magnitude estimate of an expected flux based on a potential horizon
formation at $r_s$ based on the Stefan-Boltzmann law. This estimate tells us
to expect an energy flux of the order
\[ \mathcal{h}_{, t} \sim \frac{N_c}{3840\pi r_s^2} \]
and a corresponding flux in $\mathcal{p}$ according to (\ref{conteq}).
Plugging in our $r_s \sim 3$ we find that in our current setup this is several
orders of magnitude smaller than the peak value of even the vacuum part $\mathcal{h}^0_f$ alone
and poses a formidable challenge for future studies.

\subsection{Computational aspects}

All numerical results presented in this paper have been obtained with a
standard Fortran 2008 code on a small number of standard PCs. A fully
independent Julia code was also written to crosscheck the main results. The
most time consuming part of the algorithm is of course the SVD, especially in
the implicit update step where it has to be invoked repeatedly. We have used
standard LAPACK routines in double precision for this task and not exploited
the fact that consecutive SVDs are of very similar matrices. We also have not
used any form of optimization or parallelization of our code beyond standard
compiler flags. Keeping this in mind, the time evolution of our default case
scenario to $t = 20$ took about 6 hours of computer time on a single core of a
standard PC. Although the SVD scales with the third power of the system size
and $\Delta t$ probably will have to be reduced for finer discretization, it
should be possible to simulate substantially larger systems on parallel
supercomputers.

\section{Conclusions and outlook}\label{conc}

\begin{figure}[th!]
  \begin{center}
    \includegraphics[width=0.8\textwidth]{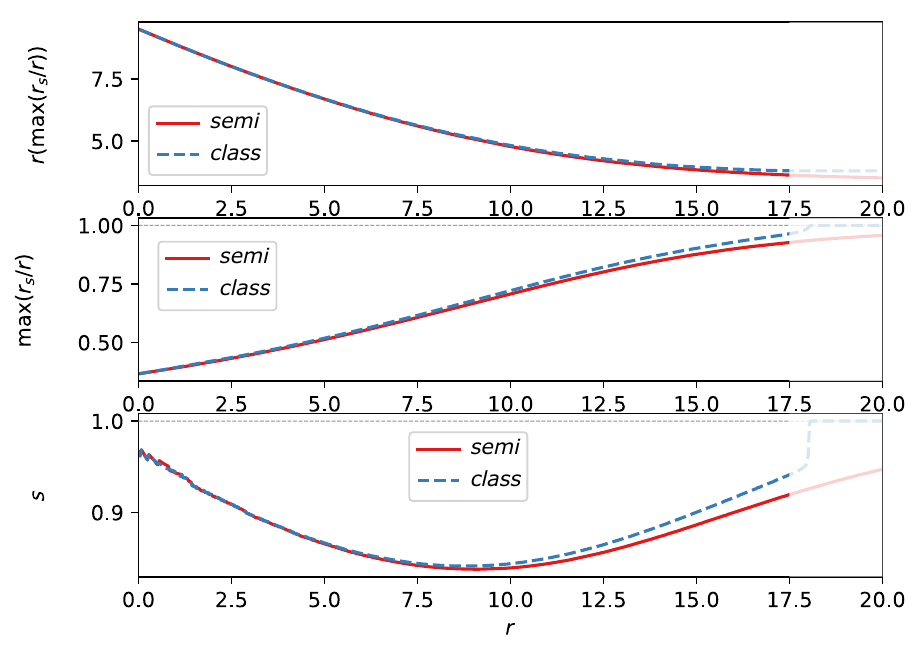}
  \end{center}
  \caption{\label{ffin}Summary of the collapse for our default case with the
  finest radial discretization $N_r~=~1600$. The top panel shows the Position
  of the maximum of $r_s / r$ vs. $t$ where $r_s$ is the local effective
  Schwarzschild radius of the metric. The middle panels shows the corresponding
  maximum value of $r_s / r$ and the bottom panel shows the mode separation parameter
  $s$ (\ref{defms}). The shaded region for large $t$ corresponds to
  $t>t_{\tmop{safe}}$.
  We can clearly see that quantum effects enhance the
  potential horizon formation and move it radially outward. We also see that
  in the semiclassical case our algorithm ultimately produces a horizon, which
  however happens in the region that we deem to be dominated by discretization
  artefacts.}
\end{figure}

In this paper we have presented a formalism to compute numerically the
semiclassical gravitational collapse of a scalar quantum field in an initial
coherent state in the angular momentum $l=0$
approximation. Fig.~\ref{ffin} summarises our main result, which is
the enhancement and radial outward shift of the peak of the energy
density by quantum effects.

As a next step, we would like to investigate how the vacuum modes
of the scalar field at higher angular momentum, which we have ignored
in this first study, will change the behaviour of the system. The
formalism we have presented here can be generalised to take these
modes into account \cite{Guenther:2021wkd}, and a detailed numerical
study is planned for a forthcoming publication.

On a more technical note, it would be nice to explore the possibility
of inhomogeneous discretizations that would increase the radial
resolution in the relevant region without too much computational
overhead. In addition and possibly related is the question of finding
a bump function that is optimized to have minimal overlap with the
singular vectors of high $\omega$ on the specific metric.

Another direction we would like to explore in the future is the use of a
different metric parametrisation, e.g. with radially infalling coordinates. An
even more ambitious goal would be to span the state of the scalar field in the
Fock space basis of the final time, which would give us access to the outgoing
particle content.

However, the ultimate question is whether the time evolution of the
semiclassical collapse is unitary and if so, whether our formalism 
could in principle trace it from the
initial infall through the (almost) formation of a horizon to the eventual
evaporation via Hawking radiation in a situation where the corresponding
classical theory does produce a horizon. Needless to say that the
technical obstacles are huge.

Finally, a problem that we have completely ignored until now is whether
semiclassical gravity is even applicable in the regime we are working in (the
effective Schwarzschild radius of our entire system viewed from the outside is
only $\sim 3.5$ Planck lengths). One direction for future investigations would
therefore be to study larger systems with a higher scalar field content so
that we are further in the semiclassical regime and check if our physical
conclusions hold there, too.

\begin{acknowledgments}
C.H. would like to thank Stephan D\"urr for helpful discussions on
the numerical aspects of this work.  This work was in part supported
by the DFG grant SFB-TR 55 and by the Excellence Initiative of
Aix-Marseille University - A*MIDEX (ANR-11-IDEX-0001-02), a French
“Investissements d’Avenir” program, through the Chaire d’Excellence
program.
\end{acknowledgments}

\clearpage

\bibliography{collapse}{}

\end{document}